\documentclass[aps,prd,nofootinbib,showkeys,preprint,floatfix]{revtex4}

\usepackage{amssymb}
\usepackage{amsmath}
\usepackage{amsfonts}
\usepackage{graphicx}
\usepackage{color}
\usepackage{xspace}
\usepackage{ulem}
\usepackage{lscape}
\usepackage{ wasysym }

\usepackage{multirow,rotating}

\def\gsim{\raise0.3ex\hbox{$\;>$\kern-0.75em\raise-1.1ex\hbox{$\sim\;$}}}
\def\lsim{\raise0.3ex\hbox{$\;<$\kern-0.75em\raise-1.1ex\hbox{$\sim\;$}}}

\def\znbb{0\nu\beta\beta}
\def\meff{\langle m_{\nu} \rangle}
\def\rpv{R_P \hspace{-1.2em}/\;\:}

\def\g{g \hspace{-0.5em}/}

\newcommand{\ba}[1]{\begin{eqnarray} \label{(#1)}}
\newcommand{\ea}{\end{eqnarray}}

\newcommand{\AddrAHEP}{
  {\it AHEP Group, Instituto de F\'{\i}sica Corpuscular --
    C.S.I.C./Universitat de Val{\`e}ncia \\
    Edificio de Institutos de Paterna, Apartado 22085,
  E--46071 Val{\`e}ncia, Spain}}

\newcommand{\AddrDO}{%
Fakult\"at f\"ur Physik, Technische Universit\"at Dortmund, 
D-44221, Dortmund, Germany}

\newcommand{\AddrUFSM}{
Universidad T\'ecnica Federico Santa Mar\'\i a, \\ 
Centro-Cient\'\i fico-Tecnol\'{o}gico de Valpara\'\i so, \\ 
Casilla 110-V, Valpara\'\i so,  Chile}

\def\gsim{\raise0.3ex\hbox{$\;>$\kern-0.75em\raise-1.1ex\hbox{$\sim\;$}}}
\def\lsim{\raise0.3ex\hbox{$\;<$\kern-0.75em\raise-1.1ex\hbox{$\sim\;$}}}

\begin{document}

\preprint{IFIC/13-34}  

\title{Short-range mechanisms of neutrinoless double beta decay at the LHC}

\author{J.C. Helo} \email{juan.heloherrera@gmail.com}\affiliation{\AddrUFSM}
\author{M. Hirsch} \email{mahirsch@ific.uv.es}\affiliation{\AddrAHEP}
\author{H. P\"as}\email{heinrich.paes@uni-dortmund.de}\affiliation{\AddrDO}
\author{S.G. Kovalenko}\email{Sergey.Kovalenko@usm.cl}\affiliation{\AddrUFSM}



\begin{abstract}
Lepton number violation (LNV) mediated by short range operators can
manifest itself in both neutrinoless double beta decay ($\znbb$) and
in processes with same sign dilepton final states at the LHC.  We
derive limits from existing LHC data at $\sqrt{s}=8$ TeV and compare
the discovery potential of the forthcoming $\sqrt{s}=14$ TeV phase of
the LHC with the sensitivity of current and future $\znbb$ decay
experiments, assuming the short-range part of the $\znbb$ decay
amplitude dominates.  We focus on the first of two possible topologies
triggered by one fermion and two bosons in the intermediate state.  In
all cases, except for the pure leptoquark mechanism, the LHC will be
more sensitive than $\znbb$ decay in the future. In addition, we
propose to search for a charge asymmetry in the final state leptons
and to use different invariant mass peaks as a possibility to discriminate 
the various possible mechanisms for LNV signals at the LHC.

\end{abstract}

\maketitle

\section{Introduction}
\label{Sec:Int}

Neutrinoless double beta ($\znbb$) is well-known as a sensitive probe
for lepton number violating (LNV) extensions of the standard model
(SM).  Possible contributions to the $\znbb$ decay amplitude beyond
the minimal mass mechanism \footnote{Exchange of a Majorana neutrino
  between two SM charged-current vertices leads to an amplitude ${\cal
    A}^{\znbb} \propto \meff \equiv \sum_i U_{ei}^2 m_i$, the
  so-called mass mechanism.} have been discussed in the literature for
many models: Left-right (LR) symmetric extensions of the SM
\cite{Mohapatra:1980yp,Doi:1985dx}, R-parity violating supersymmetry,
both trilinear $\rpv$ \cite{Mohapatra:1986su,Hirsch:1995zi} and
bilinear $\rpv$ \cite{Faessler:1997db,Hirsch:1998kc}, leptoquarks
\cite{Hirsch:1996ye}, sterile neutrinos
\cite{Bamert:1994qh,Benes:2005hn}, composite neutrinos
\cite{Panella:1994nh}, Kaluza-Klein towers of neutrinos in models with
extra dimensions \cite{Bhattacharyya:2002vf}, colour octet scalars
\cite{Choubey:2012ux} or colour sextet diquarks
\cite{Brahmachari:2002xc,Gu:2011ak,Kohda:2012sr}. A recent review of
``exotics'' in $\znbb$ decay can be found, for example, in
\cite{Deppisch:2012nb}.

However, an observation of $\znbb$ decay will not easily be
interpreted as evidence for any specific model. Several ideas to
distinguish different contributions to $\znbb$ decay have been
discussed in the literature, among them are: (i) Measure the angular
distribution of the outgoing electrons
\cite{Doi:1985dx,Arnold:2010tu}; (ii) Compare rates in
$0\nu\beta^+/EC$ decays with $0\nu\beta^-\beta^-$ decays
\cite{Hirsch:1994es} and (iii) compare rates of $0\nu\beta^-\beta^-$
decays in different nuclei \cite{Deppisch:2006hb}. In principle, all
these three methods could serve to distinguish the long-range
right-handed current term (denoted $\epsilon^{V+A}_{V+A}$ in our
notation and $\langle\lambda\rangle$ in the notation of
\cite{Doi:1985dx}) from other contributions.  However, distinguishing
among all the remaining contributions by measurements from $\znbb$
decay experiments only seems practically impossible, mainly due to the
large uncertainties in the nuclear matrix element calculations.

Contributions to the $\znbb$ decay rate can be divided into a
long-range~\cite{Pas:1999fc} and a short-range~\cite{Pas:2000vn} part.
In long-range contributions a light neutrino is exchanged between two
point-like vertices, not necessarily SM charged-current vertices.
This can lead, in those cases where one of the vertices contains a
violation of $L$ by $\Delta L=2$, to very stringent limits on the new
physics scale $\Lambda/(\lambda_{\mathrm{eff}}^{\rm LNV})\gsim$
($100-1000$) TeV. Here, $\lambda_{\mathrm{eff}}^{\rm LNV}$ is some
effective lepton number violating (LNV) coupling depending in the
model under consideration. In the short-range part of the amplitude,
on the other hand, all exchanged particles are heavy. \footnote{From
the view-point of $\znbb$ decay, heavy means masses greater than
(only) a few GeV, since the scale to compare with is the nuclear Fermi
scale, $p_F\simeq (100-200)$ MeV.} $\znbb$ decay in this case behaves
as a true effective dimension-9 operator:
\begin{equation}\label{eq:effop}
\mathcal{O}^{\znbb}_{d=9} = \frac{c_9}{\Lambda^5}
{\bar u}{\bar u}d d {\bar e}{\bar e}.
\end{equation}
The general decomposition of $\mathcal{O}^{\znbb}_{d=9}$ has very
recently been given in \cite{Bonnet:2012kh}. Using the results of
\cite{Pas:2000vn} and \cite{Bonnet:2012kh}, one finds that current
limits on the $\znbb$ decay half-lives for $^{76}$Ge
\cite{KlapdorKleingrothaus:2000sn} and $^{136}$Xe
\cite{Auger:2012ar,Gando:2012zm}, both of the order of
$T_{1/2}^{\znbb}\gsim 10^{25}$ ys, correspond to roughly $\Lambda
\gsim (1.2-3.2) g_{eff}^{4/5}$ TeV, where $g_{eff}$ is some effective
coupling (see section \ref{Sec:Pheno}) depending on the exact
decomposition.  Obviously, new physics at such scales should be
testable at the LHC.

\begin{figure}[tbh]
\hskip-10mm\includegraphics[width=0.35\linewidth]{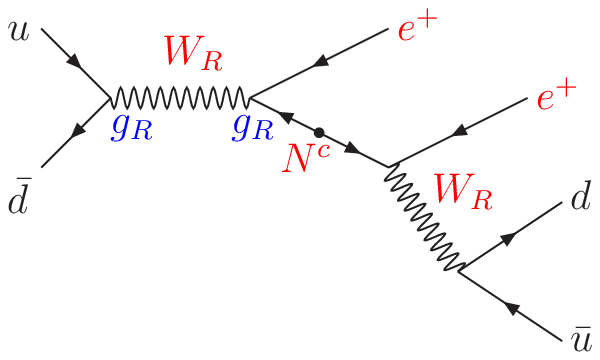}
\hskip10mm\includegraphics[width=0.5\linewidth]{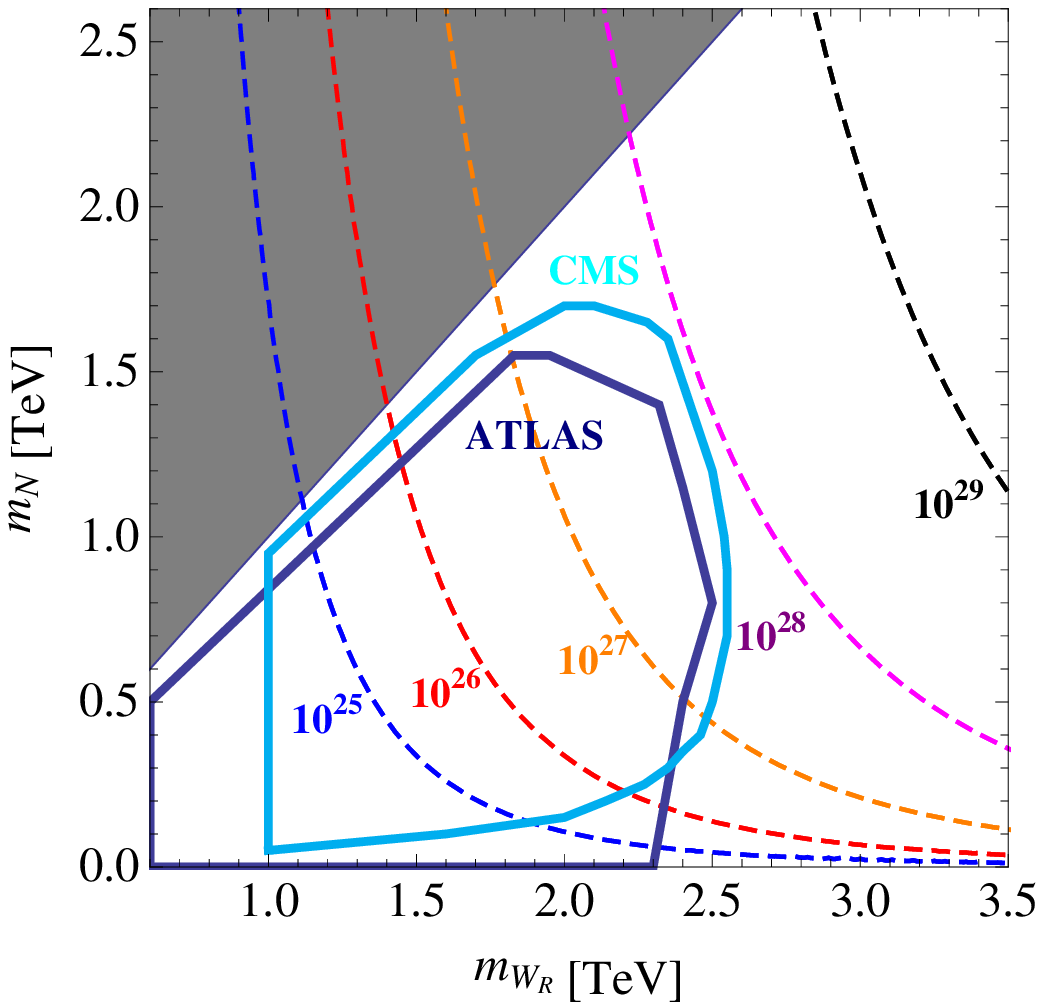}
\caption{Example diagram (left) and comparison of sensitivities of the LHC
experiments with $\znbb$ decay (right) for a (manifest) left-right symmetric
extension of the standard model. Shown are the expected contours for
the $\znbb$ decay half-life of $^{76}$Ge (contours for $^{136}$Xe are
very similar) in comparison with the excluded regions from two recent
experimental studies by ATLAS \cite{ATLAS:2012ak} and CMS
\cite{CMS:PAS-EXO-12-017} in the plane $m_{N}-m_{W_R}$, see text.} 
\label{fig:LHCLR}
\end{figure}

The prototype example of a short-range contribution which has been
discussed both for double beta decay and at the LHC are diagrams
mediated by right-handed W-bosons arising in left-right-symmetric
extension of the Standard Model
\cite{Mohapatra:1980yp,Keung:1983uu}. Here, $W_R$ can be produced
resonantly on-shell at the LHC and will decay to a right-handed
neutrino plus charged lepton, see fig.  (\ref{fig:LHCLR}). The
right-handed neutrino decays via an off-shell $(W_R)^*$, thus the
signal is both, like-sign and opposite sign, dileptons with (at least)
two jets and no missing energy
\cite{Keung:1983uu,Ferrari:2000sp,Gninenko:2006br,Bansal:2009jx}.
Fig. (\ref{fig:LHCLR}) shows a comparison of sensitivities of the LHC
experiments with $\znbb$ decay within this framework.  Contours show
the expected half-lifes for $^{76}$Ge $\znbb$ decay using the nuclear
matrix elements of \cite{Hirsch:1996qw}.  Note that contours for the
$\znbb$ decay of $^{136}$Xe are very similar.  Also shown are the
excluded regions from two recent experimental studies by ATLAS
\cite{ATLAS:2012ak} and CMS \cite{CMS:PAS-EXO-12-017}. Both LHC
analyses assume that the coupling of the right-handed $W$ boson to
fermions has exactly the same strength as the SM $W$ boson coupling to
fermions (``manifest left-right symmetry'') and the $\znbb$ decay
half-lives have therefore been calculated with the same value of the
coupling. However, the $\znbb$ decay rate sums over all mass
eigenstates $m_{N_i}$, which couple to electrons, while the LHC
experiments assume that $m_N$ appears on-shell in the $W_R$
decay. Thus, this comparison is strictly valid only if (i) only one
heavy neutrino appears in the LHC decay chain and (ii) this neutrino
couples only to electron pairs (i.e. possible generation mixing is
neglected for simplicity here). Note that limits on $W_R$ combining
the electron and muon channels at the LHC are slightly more stringent
than the ones shown in fig.(\ref{fig:LHCLR})
\cite{CMS:PAS-EXO-12-017}.

We mention in passing that resonant slepton production in R-Parity
violating SUSY leads to the same like-sign dilepton signal
\cite{Dreiner:2000vf}.  The connection of $\rpv$ SUSY at the LHC with
double beta decay has been studied in 
\cite{Allanach:2009iv,Allanach:2009xx}. 
Also a variant of the diagram in 
fig. \ref{fig:LHCLR}, but with the SM $W_{L}$  bosons and a heavy sterile 
neutrino $N$ mixed with the active ones represent a mechanism of 
$\znbb$ decay, which implications for the LHC have been studied in 
\cite{Kovalenko:2009td}.

In this paper we will generalize this comparison between double beta
decay and LHC to the complete list of short-range decompositions
(``diagrams'') of topology-I (see next section) worked out in
\cite{Bonnet:2012kh}. We consider singly charged scalar bosons,
leptoquarks and diquarks, as well as the coloured fermions, which
appear in the general decomposition of the neutrinoless double beta
operator.  We mention in passing that a brief summary of our main
results has been presented before in a conference \cite{Varzielas:2012as}
and in \cite{Helo:2013dla}.

The rest of this paper is organized as follows: In section
\ref{Sec:Dec} we briefly review the general decomposition of
$\mathcal{O}^{\znbb}_{d=9}$ developed in \cite{Bonnet:2012kh} and make
contact with the Lorentz-invariant parametrization of the decay rate
worked out in \cite{Pas:2000vn}.  Section \ref{sect:xsect} discusses
the production cross sections for different scalars. A
numerical analysis, comparing current and future LHC sensitivities
with double beta decay, case-by-case for all possible scalar
contributions to the $\znbb$ decay rate, is then performed in section
\ref{Sec:Pheno}. We then turn to the question, whether different 
models (or decompositions) can actually be distinguished at the LHC, 
if a positive signal were found in the future. We discuss two types 
of observables, which allow to do so. First, in section \ref{SubSec:CA} 
we discuss the ``charge asymmetry'', i.e. the ratio of the number of 
positron-like to electron-like dilepton (plus jets) events. We then 
turn to the discussion of invariant mass peaks in \ref{SubSec:MP}.
A joint analysis of charge asymmetry and invariant mass peaks would 
allow to identify the dominant contribution to double beta decay.
Finally, we close with a short summary of our main results.

\section{Decomposition of the $d=9$ $0\nu\beta\beta$ decay operator} 
\label{Sec:Dec}

In order to be able to compare the sensitivities of the LHC and
$\znbb$ decay experiments input from both particle and nuclear physics
is needed.  In this section we will briefly recapitulate the main
results of two papers \cite{Pas:2000vn,Bonnet:2012kh}, which we will
use in the later parts of this work. In \cite{Bonnet:2012kh} a general
decomposition of the d=9 $\znbb$ decay operator was given. This work
allows to identify all possible contributions to $\znbb$ decay from
the particle physics point of view. Moreover, it makes contact with
the general Lorentz-invariant parametrization of the $\znbb$ deay rate
of \cite{Pas:1999fc,Pas:2000vn}. These latter papers developed a
general formalism and gave numerical values for nuclear matrix
elements, which allow to calculate the expected $\znbb$ decay
half-lives for any particle physics model.


\begin{figure}[tbh]
\hskip10mm\includegraphics[width=0.9\linewidth]{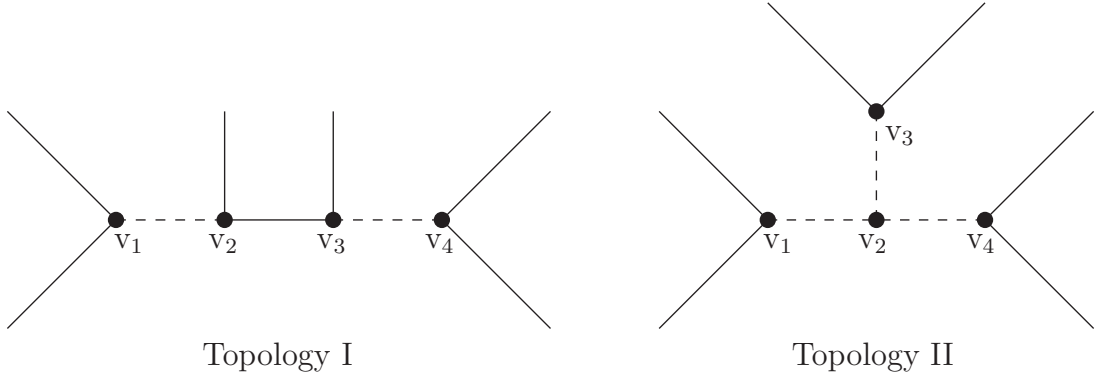}
\caption{\it \label{Fig:0nbbTopologies}The two basic tree-level
topologies realizing a $d=9$ $0\nu\beta\beta$ decay operator. External
lines are fermions; internal lines can be fermions (solid), scalars
(dashed) or vectors instead of scalars (not shown). For T-I there are
in total 3 possibilities classified as: SFS, VFS and VFV.}
\end{figure}

We start by recalling that there are only two basic topologies which
can generate the double beta decay operator at tree-level.  These are
shown in fig. (\ref{Fig:0nbbTopologies}), for brevity we will call
them T-I and T-II in the following. While all outer particles in these
diagrams are fermions, internal particles can be scalars, fermions or
vectors. For topology-I (T-I) all three possible combinations (SFS,
SFV and VFV) can lead to models, which give sizeable contributions to
$\znbb$ decay. Note that, for T-II one derivative coupling (cases VVV 
and SVV) or one dimensionful vertex (cases SSS and VVS) is needed. 
We plan to deal with T-II, which requires a slightly more complicated 
analysis, in a future publication \cite{Helo:2013xx}.

\begin{table}[h]
\begin{center}
\begin{tabular}{ccccccl}
\hline \hline 
&& 
\multicolumn{3}{c}{Mediator $(Q_{\rm em}, SU(3)_{c})$}
\\
\# & Decomposition & $S$ or  $V_{\rho}$ &  $\psi$ & $S'$ or $V'_{\rho}$ 
\\
\hline 
1-i 
&
$(\bar{u} d) (\bar{e}) (\bar{e}) (\bar{u} d)$
&
$(+1, {\bf 1}\oplus{\bf 8})$
&
$(0, {\bf 1}\oplus{\bf 8}) $
&
$(-1, {\bf 1}\oplus{\bf 8})$
\\
1-ii-a 
&
$(\bar{u} d) (\bar{u}) (d) (\bar{e} \bar{e})$
&
$ (+1, {\bf 1}\oplus{\bf 8}) $
&
$ (+ 5/3, {\bf 3})$
&
$ (+2, {\bf 1})$
\\
1-ii-b 
&
$(\bar{u} d) (d) (\bar{u}) (\bar{e} \bar{e})$
&
$(+1, {\bf 1}\oplus{\bf 8}) $
&
$(+4/3, \overline{\bf 3})$ 
&
$(+2, {\bf 1})$
\\
\hline
2-i-a
&
$(\bar{u} d) (d) (\bar{e}) (\bar{u} \bar{e})$
&
$(+1, {\bf 1}\oplus{\bf 8}) $
&
$(+4/3, \overline{\bf 3})$ 
&
$(+1/3, \overline{\bf 3})$
\\
2-i-b
&
$(\bar{u} d) (\bar{e}) (d) (\bar{u} \bar{e})$
&
$(+1, {\bf 1}\oplus{\bf 8})$
&
$ (0,{\bf 1}\oplus{\bf 8})$
&
$ (+1/3, \overline{\bf 3})$
\\
2-ii-a
&
$(\bar{u} d) (\bar{u}) (\bar{e}) (d \bar{e})$
&
$(+1, {\bf 1}\oplus{\bf 8})$
&
$ (+5/3,{\bf 3})$
&
$ (+2/3, {\bf 3})$ 
\\
2-ii-b
&
$(\bar{u} d) (\bar{e}) (\bar{u}) (d \bar{e})$
&
$(+ 1,{\bf 1}\oplus{\bf 8})$
&
$ (0,{\bf 1}\oplus{\bf 8})$
&
$ (+ 2/3, {\bf 3})$ 
\\
2-iii-a
&
$(d \bar{e}) (\bar{u}) (d) (\bar{u} \bar{e})$
&
$ (- 2/3, \overline{\bf 3})$  
&
$ (0, {\bf 1}\oplus{\bf 8})$
&
$ (+ 1/3, \overline{\bf 3})$
\\
2-iii-b
&
$(d \bar{e}) (d) (\bar{u}) (\bar{u} \bar{e})$
&
$ (- 2/3, \overline{\bf 3})$ 
&
$ (-1/3, {\bf 3_a}\oplus\overline{\bf 6_s}) $
&
$ (+ 1/3, \overline{\bf 3})$
\\
\hline
3-i
&
$(\bar{u} \bar{u}) (\bar{e})(\bar{e}) (dd)$
&
$ (+ 4/3, \overline{\bf 3}_{\bf a}\oplus{\bf 6_s}) $
&
$ (+1/3, \overline{\bf 3}_{\bf a}\oplus{\bf 6_s}) $
&
$(- 2/3, \overline{\bf 3}_{\bf a}\oplus{\bf 6_s})$  
\\
3-ii
&
$(\bar{u} \bar{u}) (d) (d) (\bar{e} \bar{e})$
&
$(+ 4/3, \overline{\bf 3}_{\bf a}\oplus {\bf 6_s}) $ 
&
$ (+5/3, {\bf 3})$  
&
$(+2, {\bf 1}) $ 
\\
3-iii
&
$(dd) (\bar{u}) (\bar{u}) (\bar{e} \bar{e})$
&
$ (+ 2/3, {\bf 3}_{\bf a}\oplus\overline{\bf 6}_{\bf s}) $  
&
$ (+4/3, \overline{\bf 3}) $ 
&
$ (+ 2, {\bf 1}) $ 
\\
\hline
4-i
&
$(d \bar{e}) (\bar{u}) (\bar{u}) (d \bar{e})$
&
$(- 2/3, \overline{\bf 3})$  
&
$( 0, {\bf 1}\oplus{\bf 8}) $ 
&
$ (+ 2/3, {\bf 3}) $  
\\
4-ii-a
&
$(\bar{u} \bar{u}) (d) (\bar{e}) (d \bar{e})$
&
$(+ 4/3,  \overline{\bf 3}_{\bf a}\oplus{\bf 6_s}) $ 
&
$ (+5/3, {\bf 3})$ 
&
$ (+ 2/3, {\bf 3}) $  
\\
4-ii-b
&
$(\bar{u} \bar{u}) (\bar{e}) (d) (d \bar{e})$
&
$ (+ 4/3,  \overline{\bf 3}_{\bf a}\oplus{\bf 6_s}) $
&
$ (+1/3, \overline{\bf 3}_{\bf a}\oplus{\bf 6_s}) $
&
$ (+ 2/3, {\bf 3}) $  
\\
\hline
5-i
&
$(\bar{u} \bar{e}) (d) (d) (\bar{u} \bar{e})$
&
$ (- 1/3, {\bf 3}) $
&
$(0, {\bf 1}\oplus{\bf 8}) $
&
$ (+ 1/3, \overline{\bf 3}) $
\\
5-ii-a
&
$(\bar{u} \bar{e}) (\bar{u}) (\bar{e}) (dd)$
&
$ (- 1/3, {\bf 3}) $
&
$ (+1/3,\overline{\bf 3}_{\bf a}\oplus{\bf 6_s}) $
&
$ (- 2/3,\overline{\bf 3}_{\bf a}\oplus{\bf 6_s}) $  
\\
5-ii-b
&
$(\bar{u} \bar{e}) (\bar{e}) (\bar{u}) (dd)$
&
$ (- 1/3, {\bf 3}) $
&
$ (-4/3, {\bf 3})$ 
&
$(- 2/3, \overline{\bf 3}_{\bf a}\oplus{\bf 6_s}) $  
\\
\hline \hline
\end{tabular}
\end{center}
\caption{\it \label{Tab:TopoI} General decomposition of the $d=9$
operator ${\bar u} {\bar u} dd{\bar e}{\bar e}$ for topology~I.  
The chirality of outer fermions is left unspecified, thus the mediators
are given with the charge $Q_{\rm em}$ of electromagnetic $U(1)_{\rm
em}$ and that of colour $SU(3)_{c}$. The symbols $S$ and $S'$ denote
scalars, $V_\rho$ and $V_\rho'$ vectors, and $\psi$ a fermion. The  
table follows the recent paper \cite{Bonnet:2012kh}, where more complete 
tables including chiralities can be found.}
\end{table}

For assigning the fermions to the outer legs in
fig. (\ref{Fig:0nbbTopologies}) for T-I there exist a total of 18
possibilities. These are listed in table (\ref{Tab:TopoI}), together
with the electric charge and possible colour transformation properties
of the intermediate state particles. Note that in these tables the
chiralities of the fermions are not given, thus the hypercharge of the
mediators is not fixed. We will come back to this point below.  
The table is valid for both, scalars and vectors, although later 
on we will concentrate on the scalar case. The results for vectors 
are very similar (apart from some minor numerical factors), so we 
will only briefly comment on these differences in our numerical 
analysis.

Table (\ref{Tab:TopoI}) contains six decompositions in which the
intermediate state fermion has zero electric charge. All T-I like
contributions to $\znbb$ decay, discussed in the literature prior to
\cite{Bonnet:2012kh}, are variants of these six decompositions.  Just
to mention two examples, T-I-1-i with vectors coupling to right-handed
fermions correspond to the $W_R-N-W_R$ exchange diagram of the
left-right symmetric model, discussed briefly in the introduction,
while the up-squark diagram of trilinear $R$-parity breaking
supersymmetry \cite{Mohapatra:1986su,Hirsch:1995zi} is classified as
SFS of T-I-4-i with chirality $({\bar u}_Lu_R^c)({\bar e}_Ld_R)({\bar
e}_Ld_R)$.  The remaining 12 decompositions all require fractionally
charged fermions with non-trivial colour transformation
properties. They require also the presence of either diquarks or
leptoquarks or both.

For most, but not all possibilities listed in table (\ref{Tab:TopoI})
two possibilities for the colour of the intermediate states
exist. This is a straight-forward consequence of the SU(3)
multiplication rules: ${\bf 3}\otimes {\bf \bar 3}= {\bf 1} + {\bf 8}$
and ${\bf 3}\otimes {\bf 3}= {\bf\bar 3}_a + {\bf 6}_{\bf s}$.  The
exception is the case of scalar diquarks, where in all cases except
2-iii-b only the ${\bf 6}_{\bf s}$ contributes, since the (scalar)
anti-triplet coupling to two identical fermions vanishes
\cite{Han:2010rf}.

\begin{figure}[htb]
\centering
\begin{tabular}{cc}
\includegraphics[width=1.0\linewidth]{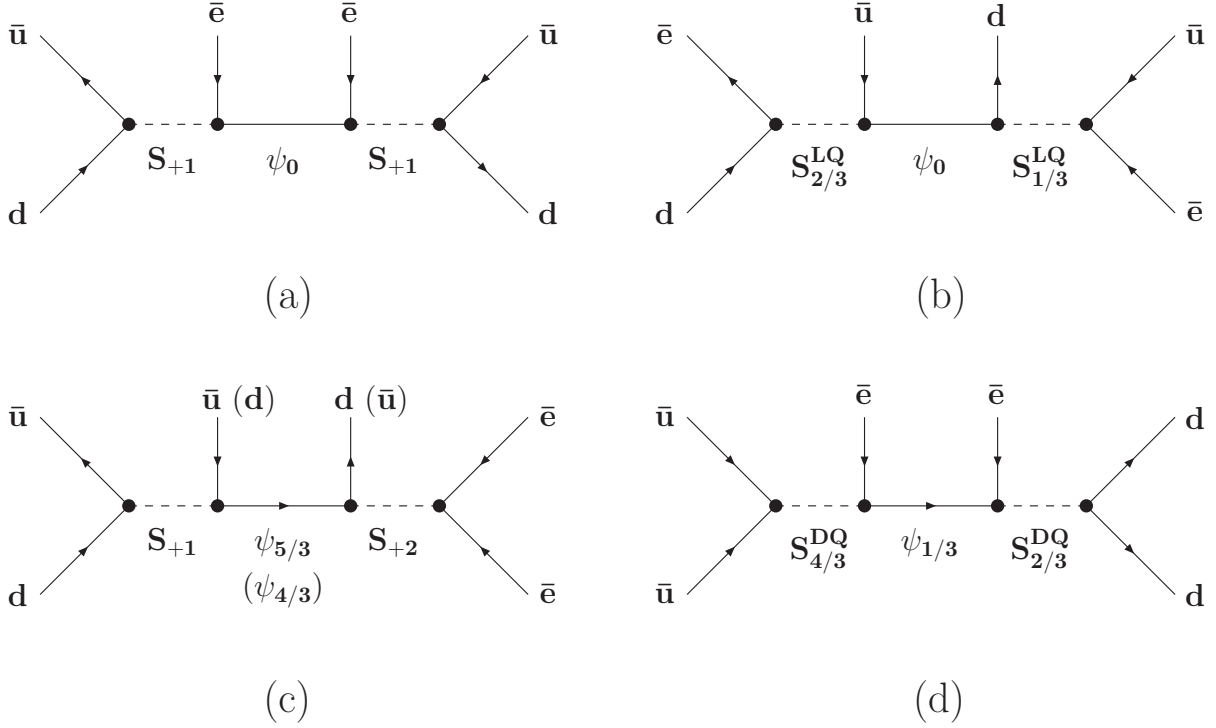}
\end{tabular}
\vskip-3mm
\caption{Example diagrams for short-range double beta decay, see text.}
\label{fig:Diags}  
\end{figure}

Fig. (\ref{fig:Diags}) shows some example diagrams, corresponding 
to the decompositions (1-i) (diagram a); (2-iii-a) (diagram b); 
(1-ii) (diagram c) and (3-i) (diagram d). These examples contain 
at least one example for each of the six different scalars and the 
four different fermions which appear in table (\ref{Tab:TopoI}). 
Diagrams for all other decompositions can be straightforwardly derived 
using the table. Note that, assigning all outer fermions to be right-handed 
and replacing $S_{+1}$ by a vector corresponds to the diagram for 
the LR-symmetric model, discussed in the introduction.

In \cite{Pas:1999fc,Pas:2000vn} a general Lorentz-invariant description 
of the $\znbb$ decay rate has been derived. The Lagrangian for the 
short-range part of the amplitude can be written as \cite{Pas:2000vn}
\begin{eqnarray}
\mathcal{L}=\frac{G_F^2}{2}m_p^{-1}\left(
            \epsilon_1 JJj+\epsilon_2 J^{\mu\nu}J_{\mu\nu}j
           +\epsilon_3 J^{\mu}J_{\mu}j+\epsilon_4 J^{\mu}J_{\mu\nu}j^{\nu}
           +\epsilon_5J^{\mu}Jj_{\mu}\right)\, .
\label{eps_short}
\end{eqnarray}
Here we omitted the chiral indices for clarity. However, for the case
of $\epsilon_3-\epsilon_5$, where chirality changes play a role in the value 
of the neutrinoless beta decay rate, the indices need to be kept.

\noindent
The low-energy energy hadronic and leptonic currents appearing 
in eq.(\ref{eps_short}) are defined as:
\begin{eqnarray}\label{eq:Currents}
J^\mu_{V\pm A}&=&\overline{u}\gamma^{\mu}(1\pm\gamma_5)d\,, \ \ J_{S\pm P}=\overline{u}(1\pm\gamma_5)d\,, \ \ 
J^{\mu\nu}=\overline{u}\sigma^{\mu\nu} d\,, 
\\ \nonumber
j^{\mu}_{A} &=& \overline{e}\gamma^{\mu} \gamma_5 e^c\,, \hspace{13mm}  j_{S\pm P} =\overline{e}(1 \pm \gamma_5)e^c\,.
\end{eqnarray}
Note that the vectorial leptonic current $j_{V}^{\mu} =
\bar{e}\gamma^{\mu} e^{c}$ is identically zero. Also the quark tensor
operator $\bar{u} \sigma^{\mu\nu} \gamma_{5} d$ is not put into the
above list since it is reducible to $J^{\mu\nu}$.

The hadronic currents in eq. (\ref{eq:Currents}) are expressed in terms 
of standard operators $({\bar u}{\cal O}d)$, adequate for the description 
of double beta decay, a low-energy process in which neutrons are converted 
into protons in a nucleus. The decompositions of table  
(\ref{Tab:TopoI}), on the other hand, are given 
in terms of quark currents. The latter can be brought into the standard 
form, eq. (\ref{eq:Currents}) by performing a Fierz transformation, 
extracting the relevant colour singlet piece(s). The corresponding 
calculations do depend on the chiralities of the outer fermions.

Once the coefficients for the basic operators of eq.(\ref{eps_short}) 
have been calculated for any given decomposition one can write the 
corresponding inverse half-life as a product of three distinct factors
\begin{equation}\label{eq:Tinv}
\left(T^{0\nu\beta\beta}_{1/2}\right)^{-1} 
    = G (\sum_i \epsilon_i {\cal M}_i)^2 .
\end{equation}
Here, $G$ is the leptonic phase space integral. Numerical values for
$G$ can be calculated accurately, see for example \cite{Doi:1985dx}.
${\cal M}_i$ are the nuclear matrix elements, they are different for
the different $\epsilon_i$. 
Their numerical values for $^{76}$Ge can be
found in \cite{Pas:2000vn}, for other isotopes see \cite{Deppisch:2012nb}. 
\footnote{For recent review of the nuclear structure theory behind the 
calculation of ${\cal M}_{i}$ see, for instance, Ref. \cite{Faessler:2012ku}
and references therein.}

\section{Cross sections}
\label{sect:xsect}

\begin{figure}[h]
\centering
\begin{tabular}{cc}
\includegraphics[width=0.5\linewidth]{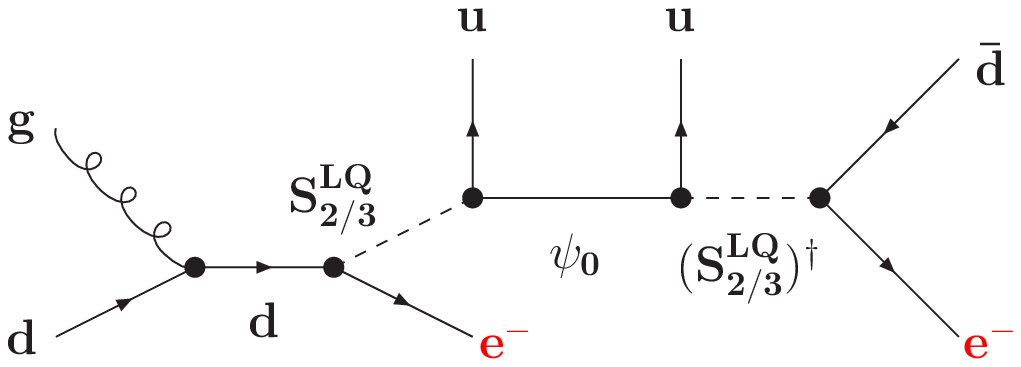}
\includegraphics[width=0.5\linewidth]{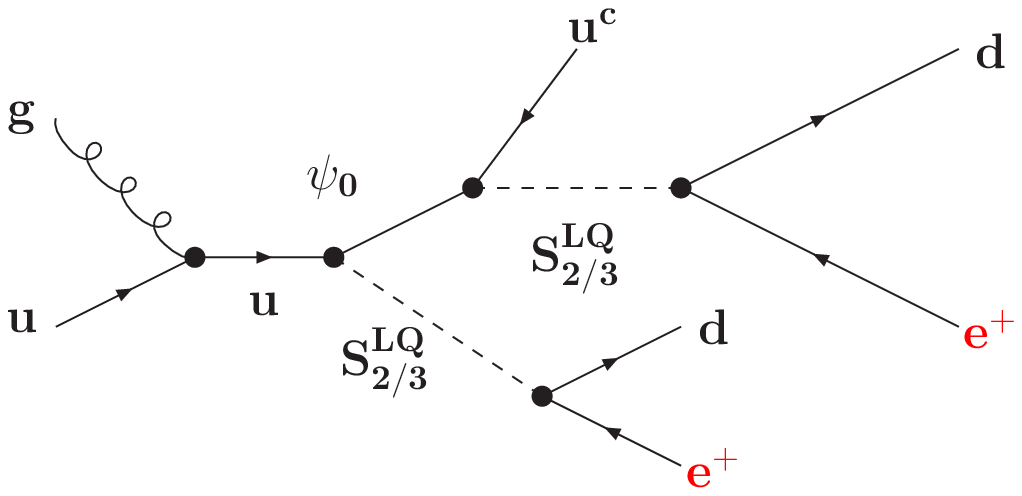} \\
\includegraphics[width=0.5\linewidth]{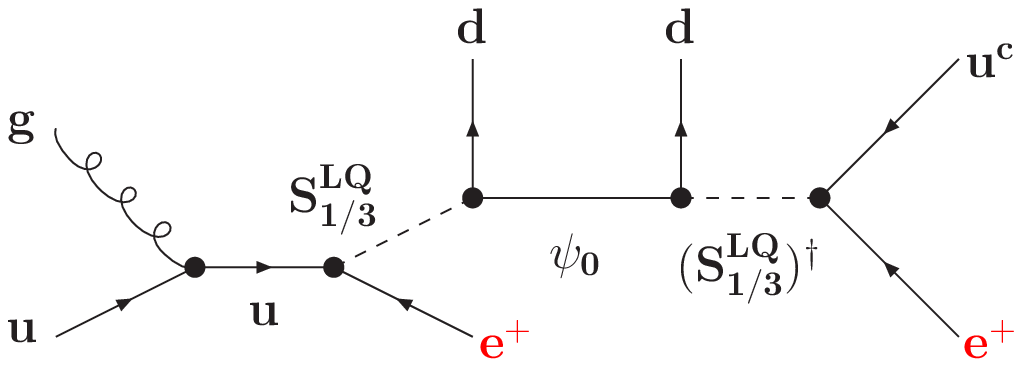}
\includegraphics[width=0.5\linewidth]{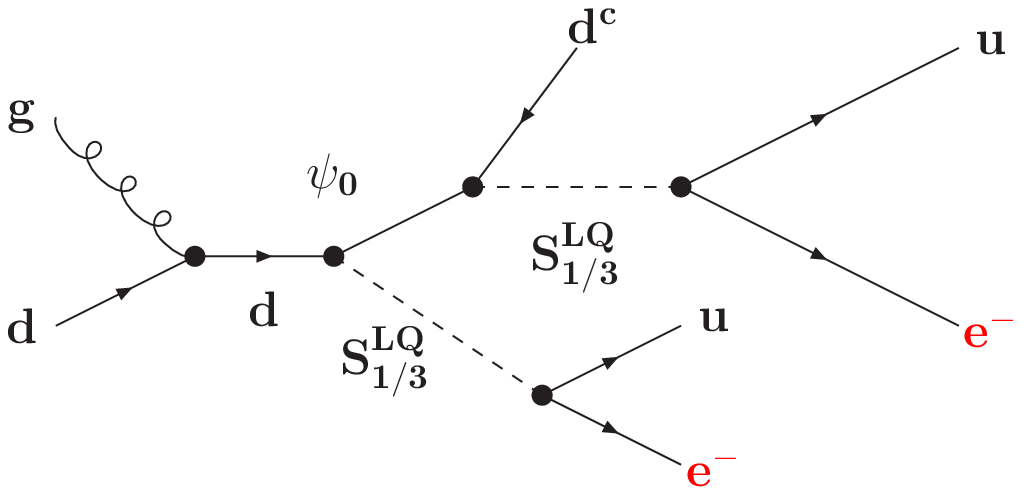} 
\end{tabular}
\vskip-3mm
\caption{Example diagrams for single leptoquark production, followed 
by LNV decay, at the LHC. For discussion see text.}
\label{fig:LQprod}
\end{figure}

In the following we discuss the production cross sections for charged
scalars ($S_{+1}$) and the two different cases each for diquarks
($S^{DQ}_{4/3}$ and $S^{DQ}_{2/3}$) and leptoquarks ($S^{LQ}_{1/3}$
and $S^{LQ}_{2/3}$) as well as their respective antiparticles.  These
five cross sections, plus the corresponding ones for vectors, are in
principle sufficient to test all 18 decompositions of the double beta
decay operator in topology-I. The $S^{DQ}_{4/3}$ occurs in
decompositions 3-i, 3-ii, 4-ii. The $S^{DQ}_{2/3}$ occurs in 3-i,
3-iii, 5-ii. The leptoquark states $S^{LQ}_{1/3}$ (and $S^{LQ}_{2/3}$)
appear in 2-i, 2-iii, and all of 5 (and 2-ii, 2-iii and all of 4,
respectively). Finally, $S_{+1}$ appears in all of 1 and in 2-i and
2-ii.  Examples of Feynman diagrams are shown in fig. \ref{fig:Diags}
and fig. \ref{fig:LQprod}. Note that the list of Feynman diagrams is
far from complete.

\begin{figure}[h]
\begin{center}
\begin{tabular}{cc}
\includegraphics[scale=0.7]{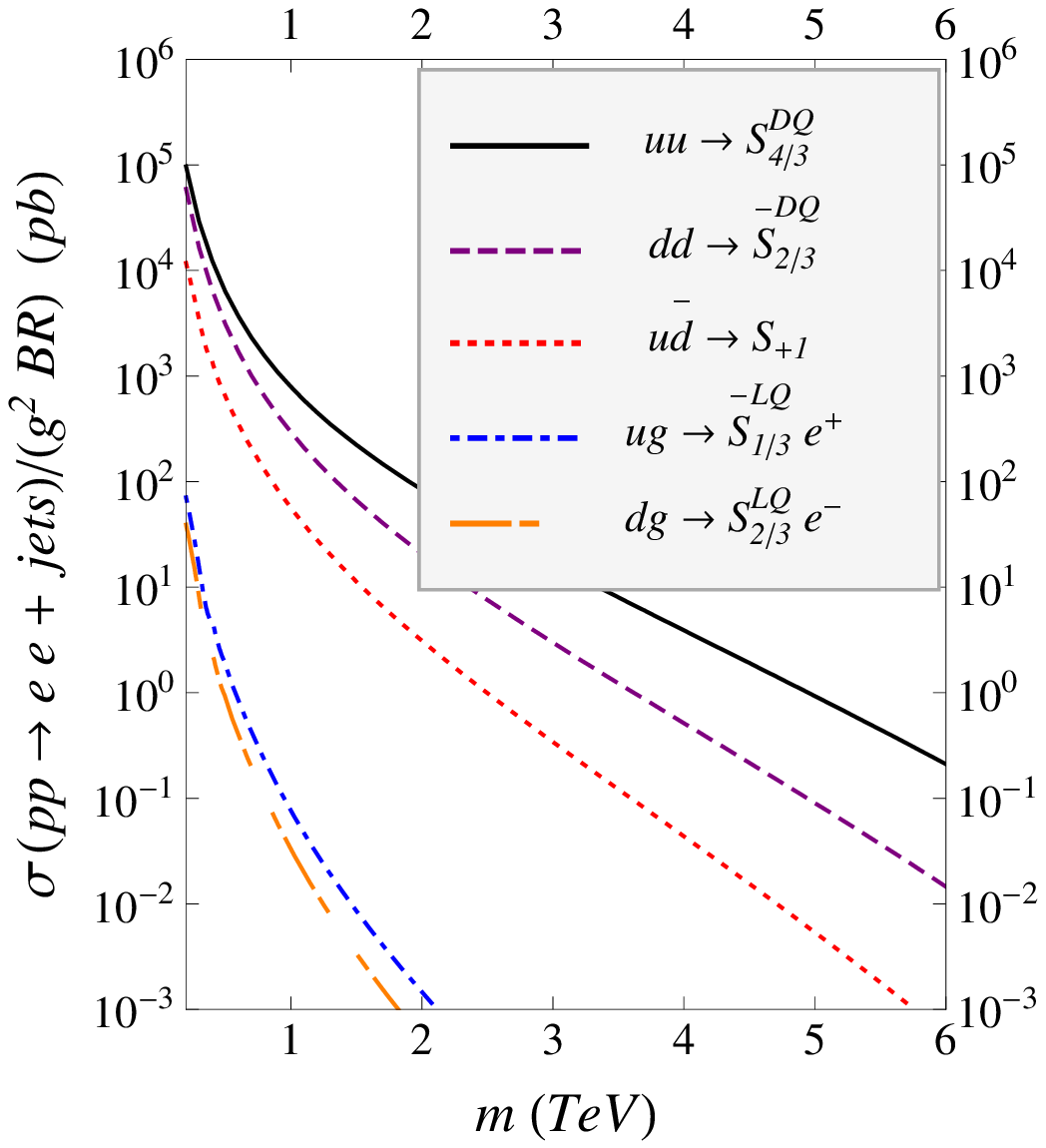}
&\includegraphics[scale=0.7]{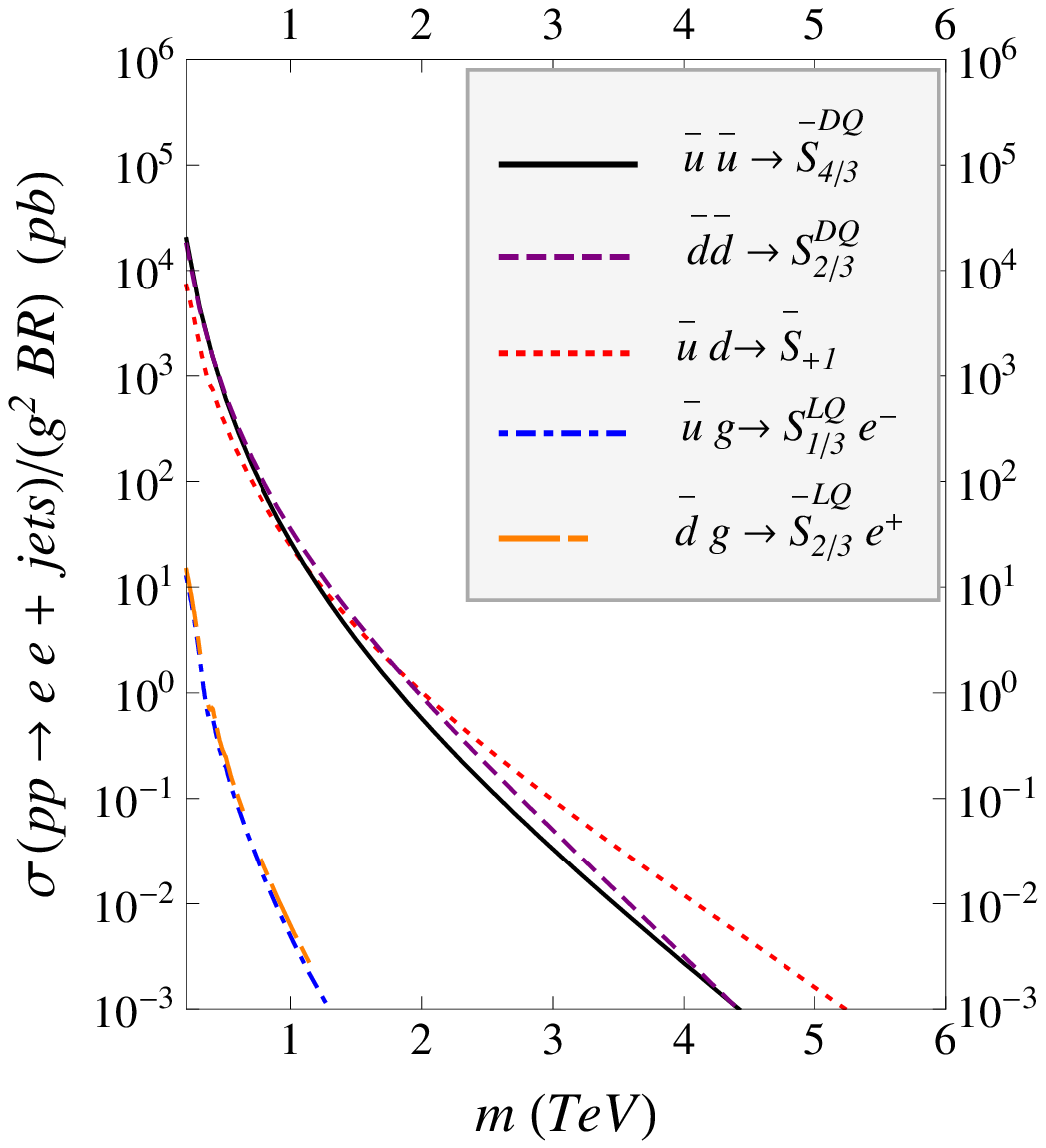}
\end{tabular}
\end{center}
\caption{\label{fig:xsectTI} Production cross sections in pb at the
LHC with $\sqrt{s}=14$ TeV for five different scalars: $S^{DQ}_{4/3}$,
$S^{DQ}_{2/3}$, $S_{+1}$, $S^{LQ}_{1/3}$ and $S^{LQ}_{2/3}$. To the
left the production of the scalar being dominantly produced (compare
the discussion of the charge asymmetry in \ref{SubSec:CA}) is being
considered. Depicted at the right is the production cross section for
its anti-particle - the scalar with the sub-dominantly produced
charge.  For $S_{+1}$, $\sigma(pp\to S_{+1})$ is only a factor
$(2-3.5$) larger than $\sigma(pp\to {\bar S_{+1}})$. For other cases
much larger ratios are found, for discussion see text.}
\end{figure}

\begin{figure}[h]
\begin{center}
\begin{tabular}{cc}
\includegraphics[scale=0.7]{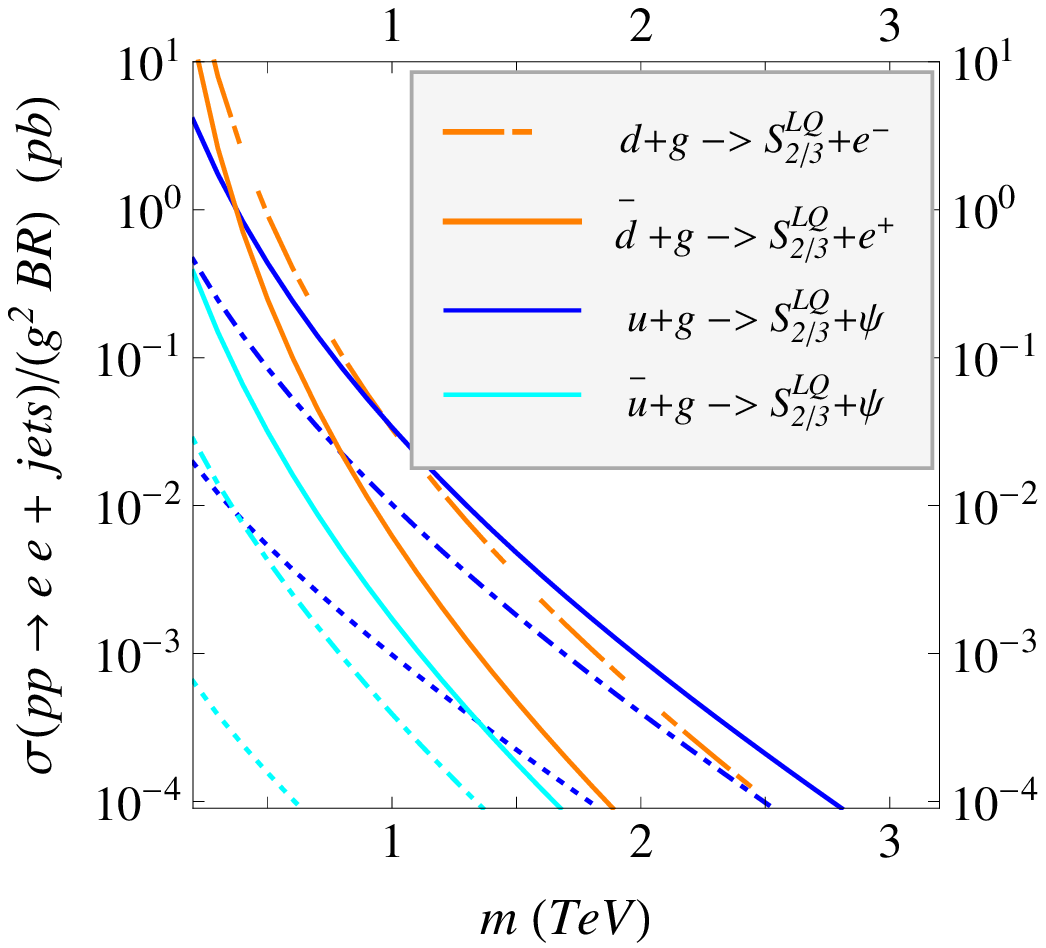}\hskip5mm
&\includegraphics[scale=0.7]{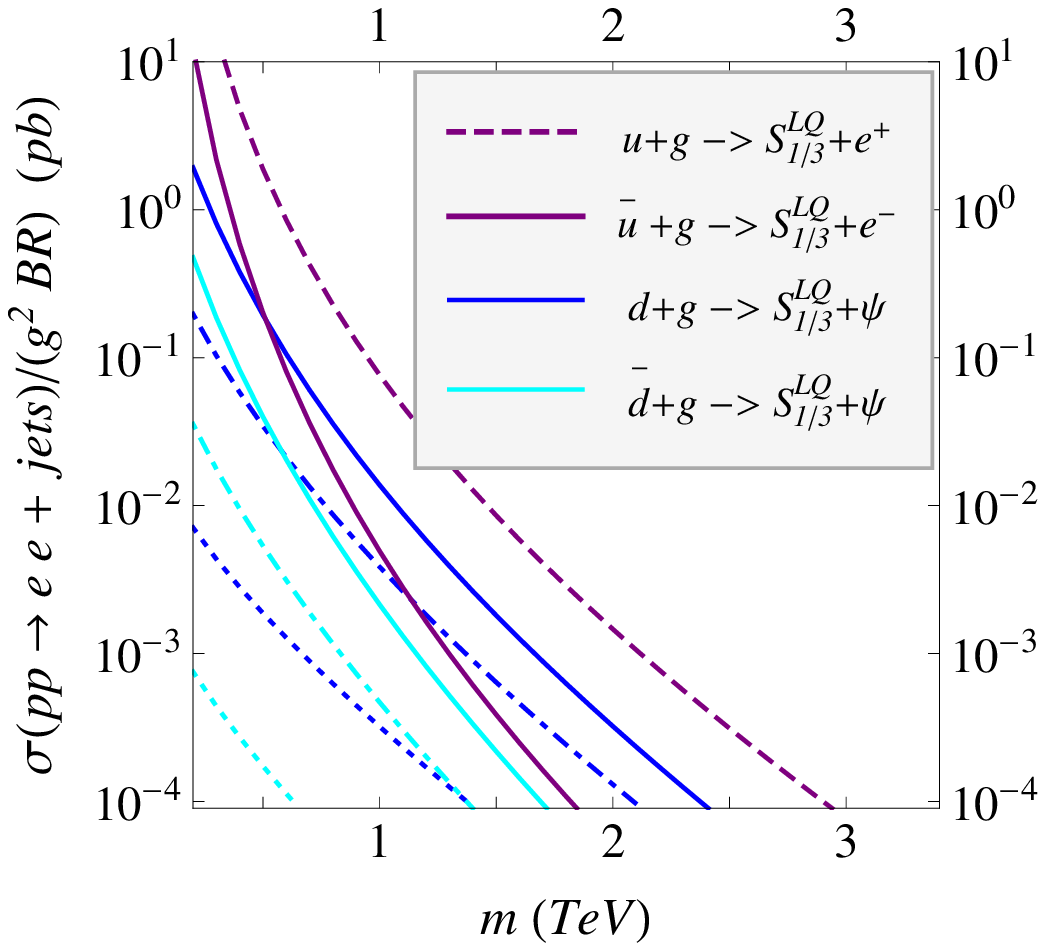}
\end{tabular}
\end{center}
\caption{\label{fig:xsectTILQ} Production cross sections in pb at the
LHC with $\sqrt{s}=14$ TeV for $S^{LQ}_{2/3}$ (left) and
$S^{LQ}_{1/3}$ (right). We show separately cross sections leading to
$e^-e^-$ and $e^+e^+$ final states. For the case of $S^{LQ}_{q}+\psi$
production, we show cross sections for three different choices of
$m_{\psi}$: $m_{\psi}=0.5$ TeV - full lines; $m_{\psi}=1.0$ TeV -
dot-dashed lines; $m_{\psi}=2.0$ TeV - dotted lines. Usually, 
$\sigma(pp\to S^{LQ}_{q}+\psi)$ is smaller than $\sigma(pp\to S^{LQ}_{q}+e)$, 
except in the case of $S^{LQ}_{2/3}$ when $m_{\psi}\ll m_{S^{LQ}_{2/3}}$.}
\end{figure}

We have implemented the corresponding Lagrangian terms given in
Appendix A into CalcHEP \cite{Pukhov:2004ca} and MadGraph5
\cite{Alwall:2011uj} for the calculation of cross sections. Example
results are displayed in figure (\ref{fig:xsectTI}) and
(\ref{fig:xsectTILQ}).

Figure (\ref{fig:xsectTI}) shows cross sections in pb for five
different scalars at LHC c.m.s. energy of $\sqrt{s}=14$ TeV. We show
$\sigma(pp \to ee + jets)/(g^2 BR)$, where $g$ stands generically for
the coupling entering the production cross section of the scalar,
``jets'' stands generically for any number of jets, and $BR$ is the
branching ratio to the final LNV state. The cross sections shown are
for colour sextets in case of diquarks, colour triplets in case of
leptoquarks. Note that for scalar diquarks coupling to the same
generation of quarks, only the sextet coupling is non-zero. For the
charged scalar, $S_{+1}$, we show the result for the colour
singlet. The cross section for a singly charged colour octet is larger
by a colour factor of $n_c=4/3$.

In fig. (\ref{fig:xsectTI}) to the left we show the ``dominant-sign'' 
production cross section, while  to the right ``wrong-sign'' charge 
production cross sections are shown. In case of $S_{+1}$, $S^{DQ}_{4/3}$ 
and $S^{LQ}_{1/3}$ the positive sign of the charge has the larger 
cross section, while for the remaining cases of $S^{DQ}_{2/3}$ and 
$S^{LQ}_{2/3}$ the negative sign is the dominant production mode. 
The ratio of dominant to subdominant cross section is, however, 
different for different scalars: For $S_{+1}$ it is in the range 
of ($2-3.5$) in the mass range shown, while for the other cases 
much larger ratios, strongly depending on the mass of the scalar 
can be found. This ``asymmetry'' in cross sections forms the basis 
of the observable ``charge asymmetry'', which we will discuss later 
in this paper.

While the charged scalar and the diquark states can be singly produced
in an s-channel resonance, as shown in fig. \ref{fig:Diags}, thus
leading to large cross sections, in case of leptoquarks the scalar LQ
is necessarily always produced in association with a lepton, see
fig. \ref{fig:LQprod}, explaining the much smaller cross sections seen
in fig. \ref{fig:xsectTI}. While the signal for diquarks (and the
charged scalar) is therefore the ``classical'' $eejj$-signal with a
mass peak in $m_{eejj}^2=m_{S_i}^2$, for LQs the signal is $ee$ with
at least three hard jets, a broader distribution in $m_{eejjj}^2$ and a
mass peak in $m_{e_2jjj}^2=m_{S^{LQ}_j}^2$, see also section \ref{SubSec:MP}. 

As shown in fig. (\ref{fig:LQprod}) leptoquarks can be produced in
association with a standard model lepton (electron/positron) or
together with one of the exotic fermions, $\psi$, see table
\ref{Tab:TopoI}. We have calculated $\sigma(pp\to S^{LQ}_{q}+\psi)$ 
for several different values of $m_{\psi}$ for both types of LQs 
(and both types of electric charge). 
In fig. (\ref{fig:xsectTILQ}) 
the results for these
cross sections are compared with 
$\sigma(pp\to S^{LQ}_{q}+e)$. 
Usually, $\sigma(pp\to S^{LQ}_{q}+\psi)$ is smaller than
$\sigma(pp\to S^{LQ}_{q}+e)$, due to the kinematical price 
of producing a heavy $\psi$ in addition to the heavy LQ. However, 
for the particular case of $S^{LQ}_{2/3}$, if $m_{\psi}\ll m_{S^{LQ}_{2/3}}$ 
the cross section for LQ plus exotic fermion production can be as 
large (or slightly larger) than $\sigma(pp\to S^{LQ}_{q}+e)$, because 
there are twice as many up-quarks in the proton than down-quarks. 

Note, that $\sigma(pp\to S^{LQ}_{2/3}+e^-)$ contributes to $e^-e^-$
type of events, while $\sigma(pp\to S^{LQ}_{2/3}+\psi)$ contributes to
$e^+e^+$ type of events. (Both contribute to $e^-e^+$ events.) For the
$S^{LQ}_{1/3}$ charges of the leptons are reversed, see
fig. (\ref{fig:LQprod}). This will be important for the charge asymmetry,
discussed in section \ref{SubSec:CA}.

The only boson appearing in the decomposition shown in table 
\ref{Tab:TopoI}, for which the single production cross section at 
the LHC is not included in fig. (\ref{fig:xsectTI}), is the doubly 
charged scalar, $S_{+2}$. Note, however, that for all decompositions 
in which $S_{+2}$ appears, the other boson in the diagram is one of the 
five, for which cross sections are shown in fig. (\ref{fig:xsectTI}). 
In fact, most of the decompositions of table \ref{Tab:TopoI} have two
different bosons in the left and right part of the diagrams and
thus, if both are within reach of the LHC, will lead to multiple
``bumps'' in the invariant mass distribution of $eejj$ (or $e_2jjj$), 
see the discussion in section \ref{SubSec:MP}.

Finally, we have also calculated the pair production cross section for
$\sigma(pp \to \psi_{1/3}{\bar \psi_{1/3}})$. For coloured fermions at
the LHC, the production cross section is dominated by gluon-gluon
fusion, thus $\sigma(pp \to \psi_{4/3}{\bar \psi_{4/3}})$ and 
also $\sigma(pp \to \psi_{5/3}{\bar \psi_{5/3}})$ have very similar 
values, while for a charge-neutral color octet cross sections are larger 
than for the case $\sigma(pp \to \psi_{1/3}{\bar \psi_{1/3}})$ by a 
corresponding colour factor. Pair production of
colored fermions provides a different signal as test for double beta
decay, since the minimal number of jets here is 4 (compared to 2 or 3
in all other cases). Cross sections are
larger than 1 fb up to masses of around 2 TeV and larger than 0.1 fb
up to 2.5 TeV. We will also briefly discuss invariant mass peaks for pair
production in section \ref{SubSec:MP}.

\section{Phenomenology}

\subsection{Status of related LHC searches}
\label{subsect:lhcstat}

Both, the ATLAS \cite{ATLAS:2012ak} and the CMS \cite{CMS:PAS-EXO-12-017} 
collaborations have published searches for events with dilepton plus 
jets (``$eejj$''). In both cases, limits on right-handed $W$-bosons 
and heavy right-handed neutrinos, motivated by the left-right symmetric 
extension of the standard model \cite{Pati:1974yy,Mohapatra:1974gc}, 
have been derived, see fig. (\ref{fig:LHCLR}). The search is based 
on the assumption that an on-shell $W_R$ is produced, decaying to 
an on-shell right-handed neutrino, i.e. $W_R \to l_1 N_l \to 
l_1 l_2 W_R^* \to l_1 l_2 jj$ \cite{Keung:1983uu}, producing two 
mass peaks in $m_{eejj}$ and $m_{e_2jj}$.

The ATLAS collaboration used $2.1$ fb$^{-1}$ of statistics at 
$\sqrt{s}=7$ TeV and searched for both, like-sign and opposite-sign, 
dileptons plus any number of jets. A number of cuts are applied to 
the data, the most important ones for us are: Leptons have to be 
isolated, with $p_T > 25$ GeV and the dilepton invariant mass 
$m_{ll}$ is required to be greater than 110 GeV. In addition, 
at least one jet has to have $p_T>20$ GeV. For larger mass differences 
betrween $W_R$ and $N$, the $N$ is significantly boosted, such that 
the two jets from the decay $N\to ljj$ are identified as a single jet. 
Such events are taken into account in the analysis and, according to 
\cite{ATLAS:2012ak}, make up up to half of the signal events. Invariant 
masses of the $m_{lljj}$ or $m_{llj}$ systems are then required to 
be larger than 400 GeV.

The main backgrounds have been identified, partially by MonteCarlo
(MC) and partially data driven, and depend on the final state
(like-sign (SS) versus opposite-sign (OS), as well as electrons versus
muons). For like-sign electrons the main background comes from ``fake
lepton events'', i.e. $W+j$, $t{\bar t}$ and QCD multi-jet production,
where one or more of the jets is misidentified as an electron. For OS
leptons, the main backgrounds are $(Z/\gamma)^*+j$ and $t{\bar t}$
events. The background for OS leptons is larger than for SS leptons by
a considerable factor ($\sim 5$), but since rough agreement between MC
and actual number of events is found in both cases the resulting upper
limits on signal cross sections are similar.

Unfortunately, \cite{ATLAS:2012ak} does not give upper limits on
$\sigma\times {\rm Br}(eejj)$ as function of $m_{eejj}$, nor does
ATLAS provide individual data sets for $e^-e^-$ and $e^+e^+$. Results
are instead presented as excluded areas in the plane ($m_N,m_{W_R}$)
for SS+OS (called ``Majorana case''), see fig. (\ref{fig:LHCLR}), and
OS-only (``Dirac case''), combining muon-type and electron-type events
and assuming $g_R=g_L$. \footnote{The classification into ``Majorana'' 
and ``Dirac'' case is done, since ATLAS assumes in its analysis that 
the fermion produced is a heavy neutrino. A Dirac neutrino will remember 
its lepton number and thus produce only electrons (positrons) in its 
decay, if the $W_R$ decayed to neutrino plus positron (electron). Thus, 
for the Dirac case only opposite sign lepton events are produced. An 
on-shell Majorana neutrino, on the other hand, will decay with 50 \% 
branching ratio into electrons and positrons each, thus producing 
both SS and OS events.}

The CMS analysis \cite{CMS:PAS-EXO-12-017} is based on 3.6 fb$^{-1}$
of data at $\sqrt{s}=8$ TeV. In their analysis, the leading lepton has
to have $p_T>60$ GeV, the subleading lepton $p_T>40$ GeV, jet
candidates $p_T>40$ GeV, as well as $m_{ll}>200$ GeV and
$m_{lljj}>600$ GeV. Events are separated into electron-like and
muon-like and separately analysed, but no charge separation within the
two sets are given, limits apply to the sum of events in the SS and OS
channels. Due to the stronger cuts on the invariant masses, absolute
background numbers in the CMS study \cite{CMS:PAS-EXO-12-017} are
similar or smaller than the corresponding background numbers in the
ATLAS study \cite{ATLAS:2012ak} despite the larger data sample.
Main backgrounds are again $(Z/\gamma)^*+j$ and $t{\bar t}$ events, 
the number of events from misidentified leptons from QCD is much 
smaller. The resulting limits in the plane ($m_N,m_{W_R}$) are 
stronger than those given by \cite{ATLAS:2012ak}, mostly due to the 
larger statistics (and also larger $\sqrt{s}$).

More important for us is that CMS presents \cite{CMS:PAS-EXO-12-017} 
also upper limits on $\sigma\times {\rm Br}(eejj)$ as function of $m_{eejj}$, 
seperately for electrons and muons. These limits assume $m_N=\frac{1}{2} 
m_{W_R}$. CMS notes that for this ratio of masses signal acceptance is 
of order (70-80) \% and drops to zero at low $m_N$, but no information 
on acceptance as function of $m_N$ is provided. Signal acceptance also 
becomes small when $m_N$ approaches $m_{W_R}$, thus for approximately 
$m_{W_R}-m_N \lsim 100$ GeV limits disappear. We will use these 
upper limits in our analysis below. However, we will assume that for 
the values of fermion masses shown, the acceptance percentage is the 
same as the one used in the plots shown by CMS. From fig.(2) of 
\cite{CMS:PAS-EXO-12-017} one can deduce that this should be a good 
approximation for fermion masses above $m_F \simeq (200-300)$ GeV. 
Note that \cite{CMS:PAS-EXO-12-017} shows cross section limits only 
for $m_{eejj} \ge 1$ TeV. For $m_{eejj}$ larger than about roughly 
$m_{eejj} \gsim 1.7$ TeV limits are of the order (2-3) fb.

In our analysis we will also use estimated sensitivity limits for the
future LHC run at $\sqrt{s}=14$ TeV. We will assume the LHC can
collect 300 fb$^{-1}$ of data. Excluding 3 signal events would then
optimistically allow to establish an upper limit on $\sigma\times {\rm
Br}(eejj)$ of $\sigma\times {\rm Br}(eejj)\lsim 0.01$ fb. Such low
values are, however, reachable only in regions of parameter space,
where background from standard model events is negligible, i.e. at the
highest values of $m_{eejj}$. For lower $m_{eejj}$, where already in
the published data a significant number of background event persists,
future data can improve limits only by much weaker factors. We do 
a simple estimate, which considers that the $t{\bar t}$ production cross 
section is about a factor $3$ higher at $\sqrt{s}=14$ TeV than at 
$\sqrt{s}=8$ TeV. Thus backgrounds should also be higher by a similar 
factor. Scaling current limits with this larger background estimate 
and taking the square root of the statistics (300 $fb^{-1}$ in the 
future, compared to roughly 3.6 $fb^{-1}$ used in \cite{CMS:PAS-EXO-12-017}, 
one can estimate the future limit for the region of $m_{eejj}$ in 
the range of ($1-2$) TeV very roughly as
$\sigma\times {\rm Br}(eejj)\lsim 0.1$ fb.

Finally, as discussed in section \ref{sect:xsect} single LQ production 
at the LHC leads to the final state $ee$ plus at least three hard 
jets. This case is only partially covered by the searches presented by 
ATLAS and CMS. In the experimental data sets the number of jets is 
required to be larger or equal to one, including in principle events with 
1, 2 and more jets. This is done simply because for large mass 
hierarchies $m_{S_i} \gg m_{\psi}$, the fermion is boosted and thus 
two jets coming from the decay of $\psi$ might be visible as a single 
jet only. On the other hand, while events with more than two hard jets 
are included in this data set, the system $ee$ plus any number of jets 
does not form a mass peak in case of LQs, as already mentioned. Peaks 
in $m_{e_2jjj}^2$, as expected for the LQs have not been searched for 
in \cite{ATLAS:2012ak,CMS:PAS-EXO-12-017}. However, one can assume 
that such a search ($ee+3j$ with peak in $m_{e_2jjj}^2$) should actually 
have similar or smaller backgrounds, due to the larger number of jets, 
than the search presented in \cite{ATLAS:2012ak,CMS:PAS-EXO-12-017}. 
In our analysis we will therefore assume also in this case that in 
the future limits of order $(0.1-1)$ fb will be reached. More precise 
numbers would require a full MonteCarlo simulation of signals and 
backgrounds, which is beyond the scope of the present work. Instead, 
see below in section \ref{Sec:Pheno}, we estimate how our limits 
will change as a function of the number of excluded events.

\subsection{Status and future of $\znbb$ limits}
\label{subsect:bbstat}

\begin{table}[h]
\begin{center}
\begin{tabular}{ccccccl}
\hline \hline 
Decomposition \# & $S-S'$ &  current limit: & future limit:
\\
\hline 
1-i, 1-ii 
&
$S_{+1}^{(1)}-S_{+1}^{(1)}/S_{+2}$
&
1.4
&
2.1
\\
\hline 
1-i, 1-ii 
&
$S_{+1}^{(8)}-S_{+1}^{(8)}/S_{+2}$
&
2.5-3.1
&
3.7-4.6
\\
\hline 
2-i, 2-ii 
&
$S_{+1}^{(1)}-S_{i}^{LQ}$
&
1.2-1.4
&
1.8-2.2
\\
\hline 
2-i, 2-ii 
&
$S_{+1}^{(8)}-S_{i}^{LQ}$
&
2.2-2.7
&
3.2-4.0
\\
\hline 
2-iii
&
$S_{i}^{LQ}-S_{j}^{LQ}$
&
1.6-3.1
&
2.4-4.6
\\
\hline 
3-i, 3-ii, 3-iii
&
$S_{i}^{DQ}-S_{j}^{DQ}/S_{+2}$
&
2.4-2.7
&
3.5-4.1
\\
\hline 
4-i, 5-i
&
$S_{i}^{LQ}-S_{i}^{LQ}$
&
2.0-2.5
&
3.0-3.7
\\
\hline 
4-ii, 5-ii
&
$S_{i}^{DQ}-S_{j}^{LQ}$
&
2.0-2.4
&
3.0-3.5
\\
\hline \hline
\end{tabular}
\end{center}
\caption{\it 
\label{Tab:bbstat} Status and future of limits on short-range operators 
from $\znbb$ decay experiments. Different decompositions result in 
different limits and depend on the helicity of the outer fermions. 
The first column gives the decompositon number, compare to table 
(\ref{Tab:TopoI}), the 2nd column indicates the exchanged scalars.
If within a certain (set of) decomposition(s) more than one operator 
can appear, depending on helicity asignments, for brevity we quote a 
range for the limit corresponding to the largest and smallest operators 
within this decomposition. ``Current limit'' are the limits assuming 
$T_{1/2}^{\znbb}(^{136}Xe) \ge 1.6 \times 10^{25}$ yr \cite{Auger:2012ar}, 
while ``future limit'' correspond to an assumed future limit of the 
order of $10^{27}$ yr. The numbers quoted are limits on $M_{eff}$ in 
TeV and scale as $g_{eff}^{(4/5)}$.}
\end{table}

As mentioned in the introduction, currently the best limits on $\znbb$
decay come from experiments on two isotopes, namely $^{76}Ge$ and
$^{136}Xe$. The Heidelberg-Moscow collaboration gives
$T^{0\nu\beta\beta}_{1/2}(^{76}{\rm Ge}) \ge 1.9 \cdot 10^{25}$ yr
\cite{KlapdorKleingrothaus:2000sn}, while the recent results from
EXO-200 and KamLAND-ZEN quote $T^{0\nu\beta\beta}_{1/2}(^{136}{\rm
Xe}) \ge 1.6 \cdot 10^{25}$ yr \cite{Auger:2012ar} and
$T^{0\nu\beta\beta}_{1/2}(^{136}{\rm Xe}) \ge 1.9 \cdot 10^{25}$ yr
\cite{Gando:2012zm}, both at the 90 \% CL. However, it is expected
that these limits will be improved within the near future. The GERDA
experiment \cite{Abt:2004yk,Ackermann:2012xja} will release first
$\znbb$ data in summer of 2013 and then move to ``phase-II'', aiming
for $T^{0\nu\beta\beta}_{1/2}(^{76}{\rm Ge})$ in excess of $10^{26}$
yr. An experiment using $^{130}Te$ in bolometers named CUORE
\cite{Alessandria:2011rc}, with senstivity order $10^{26}$ yr is
currently under construction. Proposals for ton-scale next-to-next
generation $\znbb$ experiments claim that even sensitivities in excess
$T^{0\nu\beta\beta}_{1/2} \sim 10^{27}$ yr can be reached for
$^{136}$Xe \cite{KamLANDZen:2012aa,Auty:2013:zz} and $^{76}$Ge
\cite{Abt:2004yk,Guiseppe:2011me}. For recent reviews and a list of
experimental references, see for example \cite{Barabash:1209.4241}.

In table (\ref{Tab:bbstat}) we therefore quote current and expexcted 
future limits on $M_{eff}$ from double beta decay experiments using 
$T^{0\nu\beta\beta}_{1/2}(^{136}{\rm Xe}) \ge 1.6 \cdot 10^{25}$ yr 
(current) and $10^{27}$ yr (future). Here, $M_{eff}$ and $g_{eff}$ 
are simply defined as the effective mass and couplings, which 
enter the $\znbb$ decay amplitude:
\begin{eqnarray}\label{eq:meff}
M_{eff} = (m_{S}^2m_{\psi}m_{S'}^2)^{(1/5)} \\ \nonumber
g_{eff} = (g_1g_2g_3g_4)^{(1/4)}
\end{eqnarray}
We show limits for the different decompositions assuming scalars are
exchanged. The limits on $M_{eff}$ are in TeV and scale as
$g_{eff}^{(4/5)}$. Within a given decomposition different operators
can appear in the calculation of the $\znbb$ decay half-live.  If
within a given decomposition there is more than one operator
combination that appears for the different possible helicity states,
we quote a range of limits, corresponding to the operators with the
largest and smallest possible rate within this decomposition. Numbers
are calculated using the nuclear matrix elements of
\cite{Deppisch:2012nb} and the uncertainty on $M_{eff}$ scales as
$\Delta(M_{eff}) \propto (\Delta M_{\rm Nucl.})^{(1/5)}$, where $M_{\rm Nucl}$ 
stands generically for the nuclear matrix elements. Current
limits range from $M_{eff}\gsim 1.2-3.1$ TeV, future sensitivities up
to $M_{eff}\gsim 1.8-4.6$ TeV are expected.

\subsection{LHC $\text{vs}$ $\znbb$: Numerical analysis}
\label{Sec:Pheno}

In this section we compare the LHC and $\znbb$ sensitivities of the
different decompositions for $\znbb$. In the numerical analysis we
will develop in this section we will concentrate on the case when the
fermion mass is smaller than (one of) the scalar masses.  The
numerical analysis of the case when the fermion mass is larger than
the scalar masses will be analysed in a future paper.

We can divide the discussion of all 18-decompositions into two group
of cases. The first group correspond to "symmetric " and
"like-symmetric" decompositions. The former are simply those
decompositions, in which a scalar with the same quantum numbers
appears twice in the diagram (1-i, 5-i, 4-i), while the later are
those with two different scalars but of the same kind (two different
leptoquarks or two different diquarks: 2-iii, 3-i). The second group
correspond to "asymmetric" decompositions. Those correspond to
decompositions with $S_{+2}$ and either a $S_{+1}$ or a diquark (1-ii,
3-ii, 3-iii) and decompositions with a leptoquark and either a
$S_{+1}$ or a diquark (2-i, 2-ii, 4-ii, 5-ii).

First we will derive limits from existing LHC data at $\sqrt {s} = 8$
TeV to compare then the discovery potential of the forthcoming $\sqrt
{s} = 14$ TeV phase of the LHC with the sensitivity of current and
future $0\nu \beta \beta$ decay experiments. We will begin our
discussion with the "symmetric" decomposition $(\bar u d)(\bar e
)(\bar e)(\bar u d)$.

As discussed in section IV-A the most stringent current limits from
the LHC on like-sign lepton searches come from data taken by the CMS
collaboration at $\sqrt{s} = 8$ TeV \cite{CMS:PAS-EXO-12-017}. CMS
presents also upper limits on $\sigma \times Br(eejj)$ as a function
of $m_{eejj}$. These limits apply directly to the case of the
decomposition $(\bar u d)(\bar e )(\bar e)(\bar u d)$, which describes
at LHC a produced scalar $S_{+1}$, decaying to $S_{+1}\to \psi_0 e^+$,
followed by $\psi_0 \to e^+ \bar u d$, producing two mass peaks in
$m_{eejj}$ and $m_{e_2 j j}$.

The number of eejj-like events at the LHC in general depends on a
different combination of couplings and masses than the $\znbb$ decay
amplitude. The $\znbb$ half-life depends on the effective parameters
defined in Eq. (\ref{eq:meff}) and the cross section $\sigma \times
Br(eejj)$ is, in the narrow width approximation, proportional to
$g_{udS_{+1}}^2$  and to a non trivial function $F_{S_{+1}}$   of the scalar mass
$m_{S_{+1}}$ 
We can then 
write the number of events as:
\begin{eqnarray}\label{eq:SigBrS1}
\sigma \times Br(eejj) =  \sigma(pp\to S)\times{\rm Br}(S\to eejj) =
 F_{S}\left( m_{S} \right)  g_{1}^2 \ {\rm Br}(S\to eejj),
\end{eqnarray}
defining
\begin{eqnarray}\label{FS+Def}
F_{S_{+1}}(m_{S_{+1}})=\sigma(pp\to S_{+1})/g_{udS_{+1}}^2.
\end{eqnarray}
The ${\rm Br}(S\to eejj)$ can be calculated from  
eq. (\ref{Lag-S1-psi}) and is equal to
\begin{eqnarray}
\label{BR}
Br(S \to eejj) = \frac{f(m_{\psi}/m_S) g_2^2 }
{3 g_1^2 + f(m_{\psi}/m_S) g_2^2 } \times \frac{1}{2}.
\end{eqnarray}
Here $S=S_{+1}$, $g_1 = g_{ud S_{+1}}$, $g_2 = g_{e \psi_0 S_{+1}}$, $\psi = \psi_0$ and 
$f(x)=(1-x^2)^2$. 
Note that in the limit
where all couplings are equal (and $m_{\psi_0}=0$) ${\rm Br}(S_{+1}\to
e^+e^+jj)={\rm Br}(S_{+1}\to e^+e^-jj) \simeq 1/8$. We have used
CalcHEP \cite{Pukhov:2004ca} to calculate the production cross
sections for $S_{+1}$ at the LHC. We have plotted our results in
Fig. \ref{fig:xsectTI} and compared them with the literature
\cite{Ferrari:2000sp} finding quite good agreement.

In ``symmetric''
decompositions, such as $(\bar u d)(\bar e )(\bar e)(\bar u d)$, the
effective couplings and scalar boson masses are pairwise equal,
i.e. in Eq. (\ref{eq:meff}) $g_1=g_4$, $g_2=g_3$ and $m_{S}=
m_{S^\prime}$.  Then, the effective parameters defined in
(\ref{eq:meff}) become:
\begin{eqnarray}
\label{Efective-S1}
M_{eff(S)} &=&  (m_{S}^4  m_{\psi} )^{1/5}, \ \   g_{eff(S)} = (g_{1}   g_{2} )^{1/2}.
\end{eqnarray}

\begin{figure}[h]
\includegraphics[scale=0.8]{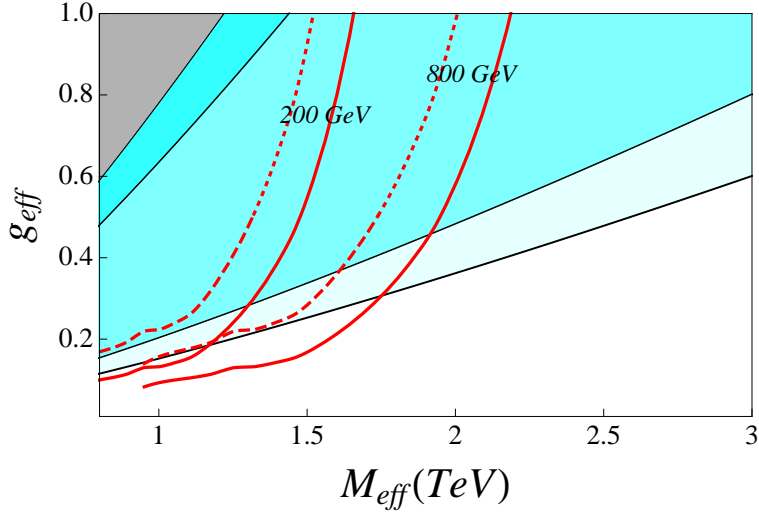}
\caption{ Current limits for the LHC at $\sqrt {s} = 8$ TeV for
  production of scalars $S_{+1}$ compared with current and future
  double beta decay experiments.  The gray region in the top left
  corner is ruled out by current $\znbb$ data. The blue region
  represents the parameter region accessible in near future $\znbb$
  experiments, whereas the red lines shows current LHC limits for
  production of scalars $S_{+1}$. Solid red lines were calculated
  using $Br(S_{+1} \to e e j j ) = 10^{-1}$ while the dashed and dotted
  red lines were calculated using $Br(S_{+1} \to e e j j ) =
  10^{-2}$ for different values of the fermion mass $m_{\psi_0} = 200$
  GeV, $\ 800$ GeV, see text.}
 \label{fig:Lim_8TeV}
\end{figure}

Eq. (\ref{eq:SigBrS1}) depends on 4 variables: the couplings $g_1$,
$g_2$ and the masses $m_{S}, m_{\psi}$.  For comparison with $\znbb$
we have expressed Eq. (\ref{eq:SigBrS1}), using
Eq. (\ref{Efective-S1}) and (\ref{BR}), in terms of 4 new variables:
the effective coupling and mass $g_{eff}, M_{eff}$, the fermion mass
$m_{\psi}$ and the $Br(S \to eejj)$. Then, using
Eq. (\ref{eq:SigBrS1}) expressed in terms of this 4 new variables, and
the current limits on $\sigma \times Br(eejj)$ presented by CMS
\cite{CMS:PAS-EXO-12-017} we can plot in the plane $g_{eff}$ versus
$M_{eff}$ bounds of the LHC for differentt values of the $Br(S_{+1}
\to e e j j )$ and the fermion mass $m_{\psi_0}$. We have drawn these
limits for $Br(S_{+1} \to e e j j ) = 10^{-1}$ (solid red lines) and
$Br(S_{+1} \to e e j j ) = 10^{-2}$ (dashed red lines) in
Fig. \ref{fig:Lim_8TeV} using different values of the fermion mass
$m_{\psi_0} = 200$ GeV, $\ 800$ GeV. For larger masses $m_{\psi_0} $
the LHC limits become more stringent except for the region
$(m_{\psi_0} - m_{S_{+1}}) \lesssim 100$ GeV, where the LHC
sensitivity becomes very small as we discussed in section
\ref{subsect:lhcstat}. Note, that the for the dotted/dashed lines, 
the part of the line, which is shown dotted correspond to values 
of $1 \le g_1=g_{u d S_{+1}} \le 2$, i.e. close to values where 
this coupling would become non-perturbative.

In addition, Fig. \ref{fig:Lim_8TeV} shows current and future limits
from $\znbb$ decay. The dark gray area is the currently excluded part
of parameter space from non-observation of $^{136}$Xe decay with
$T_{1/2}^{0\nu \beta \beta}\ge 1.6 \times 10^{25}$ ys
\cite{Auger:2012ar} and the blue area correspond to an assumed future
$0\nu\beta \beta$ decay sensitivities of $T_{1/2}^{0\nu \beta
  \beta}\ge 10^{27}$ ys.  We have used as a current limit $M_{eff}
>1.2 \ TeV \times g_{eff}^{4/5}$ and for future sensitivities up to
$M_{eff} >4.6 \ TeV \times g_{eff}^{4/5}$.  These correspond to the
most pessimistic case for the current sensitivity of $\znbb$ decay and
the most optimistic reach for $\znbb$ decay in the foreseeable future
(See Table \ref{Tab:bbstat}). As we can see from
Fig. \ref{fig:Lim_8TeV} the LHC is already competitive to $\znbb$ for
part of the parameter region of the decomposition $(\bar u d)(\bar e
)(\bar e)(\bar u d)$, especially for larger masses of
$m_{\psi_0}$. However, this mechanism is not ruled out, quite on the
contrary, most of the parameter region explored by future $\znbb$
decay experiments has not been covered yet.

Now we will analyze the discovery potential of the forthcoming
$\sqrt{s} = 14$ TeV phase of the LHC.  We will start our discussion
with the first group of decompositions, i.e. "symmetric" and
"like-symmetric" decompositions. Recall, for "symmetric"
decompositions one can use Eqs. (\ref{eq:SigBrS1})-(\ref{Efective-S1})
to describe the cross section $\sigma \times Br(eejj)$ in terms of the
effective masses and couplings relevant for $\znbb$. In the LQ case,
the LQ is produced in association with a lepton, i.e. in
Eq. (\ref{eq:SigBrS1}) we calculate $\sigma(pp\to S^{LQ}+e)\times {\rm
  Br}(S^{LQ}\to ejjj)$. For "like-symmetric" decompositions,
Eq. (\ref{Efective-S1}) is also a good approximation. This is because
both LQs or both diquarks can be produced at LHC and in turn will have
similar limits on the masses $m_{S}, m_{S^\prime}$ and couplings $g_1,
g_4$ and $g_2, g_3$.  We have used CalcHEP \cite{Pukhov:2004ca} and
MadGraph 5 \cite{Alwall:2011uj} to calculate the production cross
sections for $S_{+1}$, $S^{LQ}$, and $S^{DQ}$ at the LHC.  We have
plotted our results in Fig. \ref{fig:xsectTI} and compared them with
the literature \cite{Ferrari:2000sp,Belyaev:2005ew,Han:2010rf} and
found quite good agreement in all cases.

\begin{figure}[htbp]
\begin{minipage}[b]{.45\linewidth}
\includegraphics[width=\linewidth]{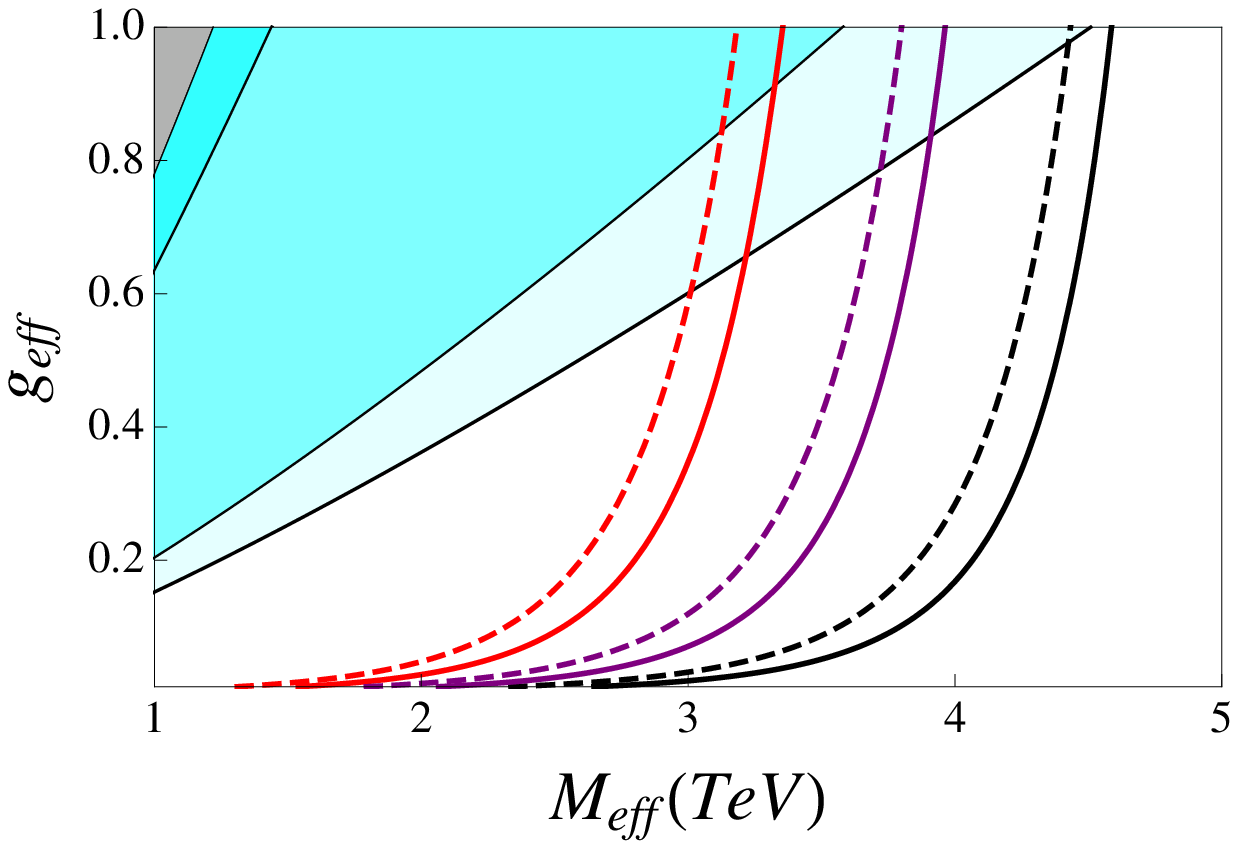}
\end{minipage}
\begin{minipage}[b]{.05\linewidth}
\hspace{1pt}
\end{minipage}
\begin{minipage}[b]{.45\linewidth}
\vspace{0pt}
\includegraphics[width=\linewidth]{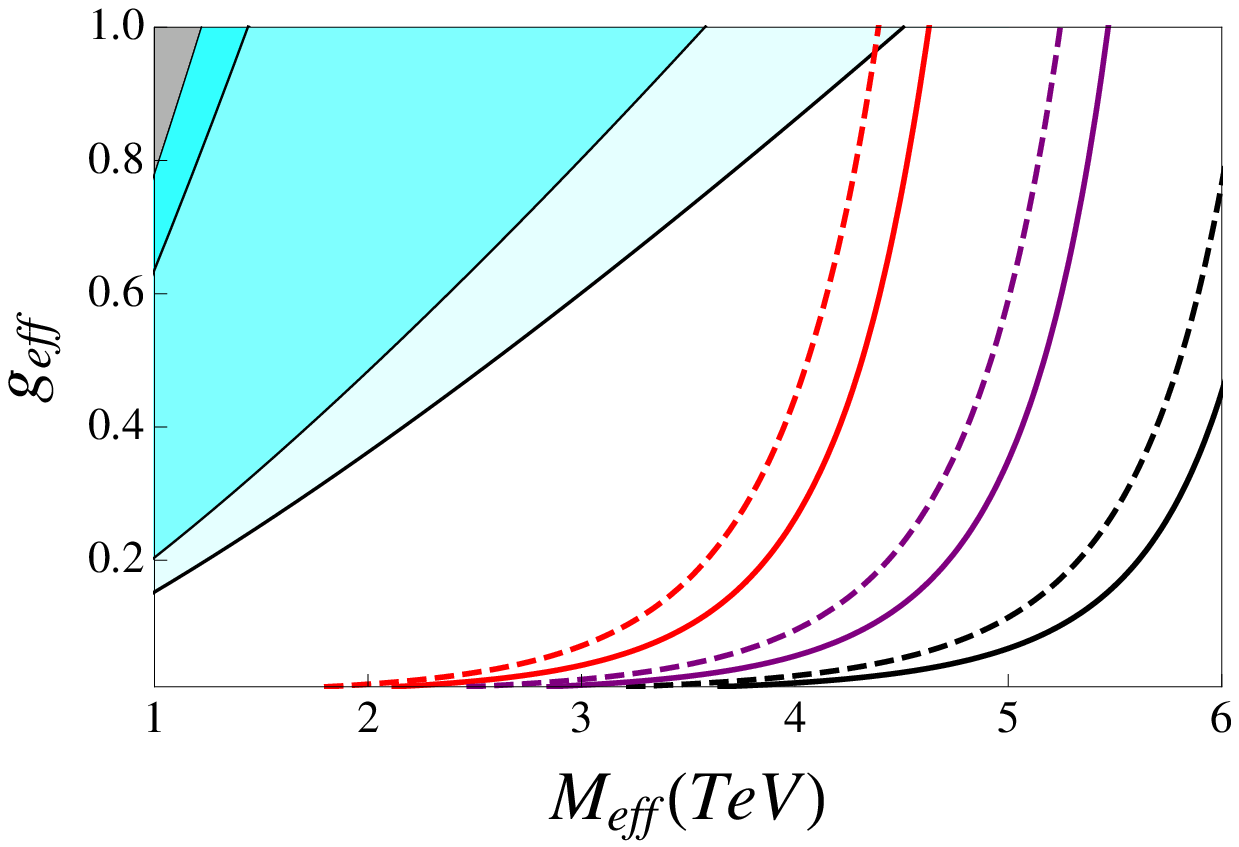}
\end{minipage}
\vspace{0.2 cm}
\caption{ Future limits for the LHC at $\sqrt {s} = 14$ TeV compared
  with current and future double beta decay experiment.  The gray
  region on the top left corner is ruled out by $\znbb$. The blue
  region represents the parametric region accessible in near future
  $\znbb$ experiments, whereas the colored lines shows sensitivity
  limits for the LHC for production of three different scalar bosons
  $S_{+1}$ (red), $S_{2/3}^{DQ}$ (purple) and $S_{4/3}^{DQ}$
  (black). Solid lines were calculated using $Br(S \to e e j j )
  = 10^{-1}$ whiles dashed lines were calculated using $Br(S \to
  e e j j ) = 10^{-2}$ for different values of the fermion mass
  $m_{\psi} = 200$  GeV  (left) and $m_{\psi} = 1000$  GeV (right).
 }
\label{fig:Lim_Sim}
\end{figure}

\begin{figure}[htbp]
\begin{minipage}[b]{.45\linewidth}
\includegraphics[width=\linewidth]{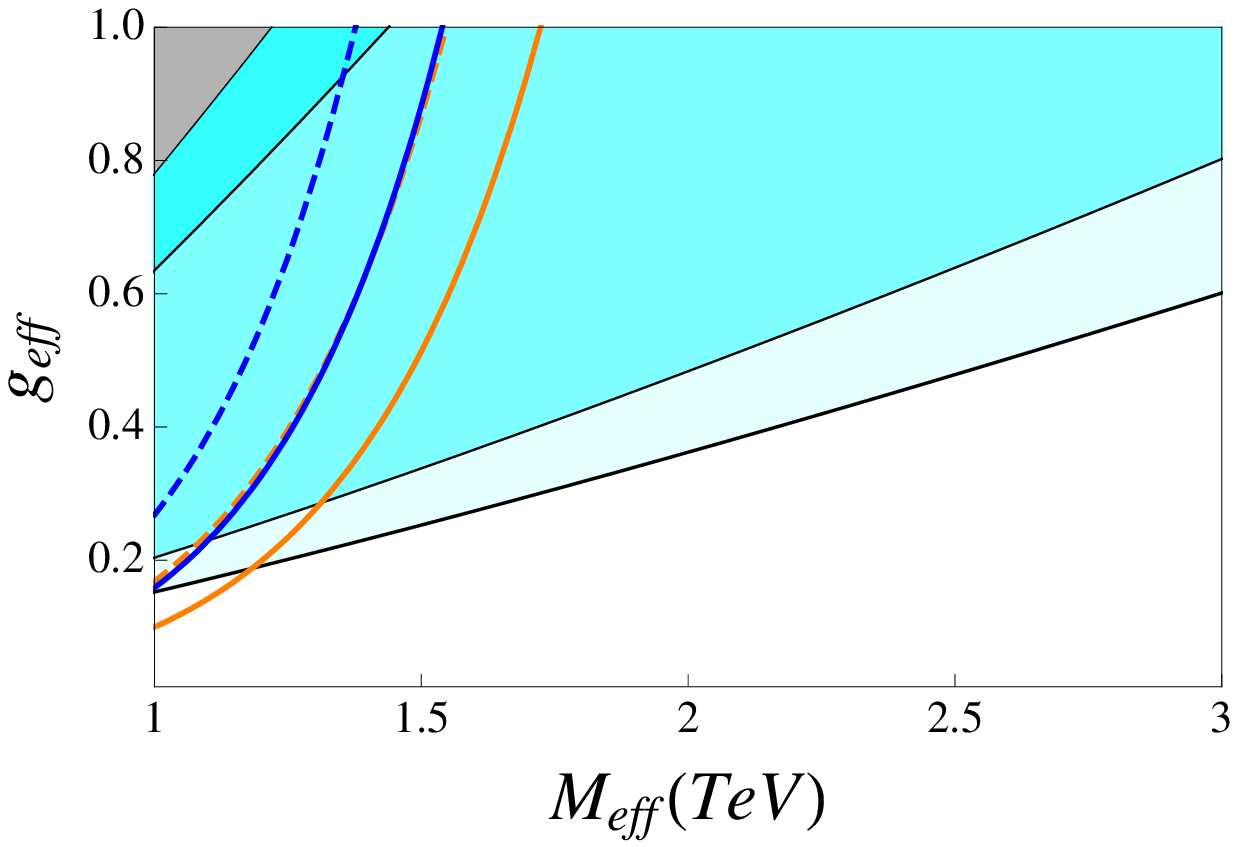}
\end{minipage}
\begin{minipage}[b]{.05\linewidth}
\hspace{1pt}
\end{minipage}
\begin{minipage}[b]{.45\linewidth}
\vspace{0pt}
\includegraphics[width=\linewidth]{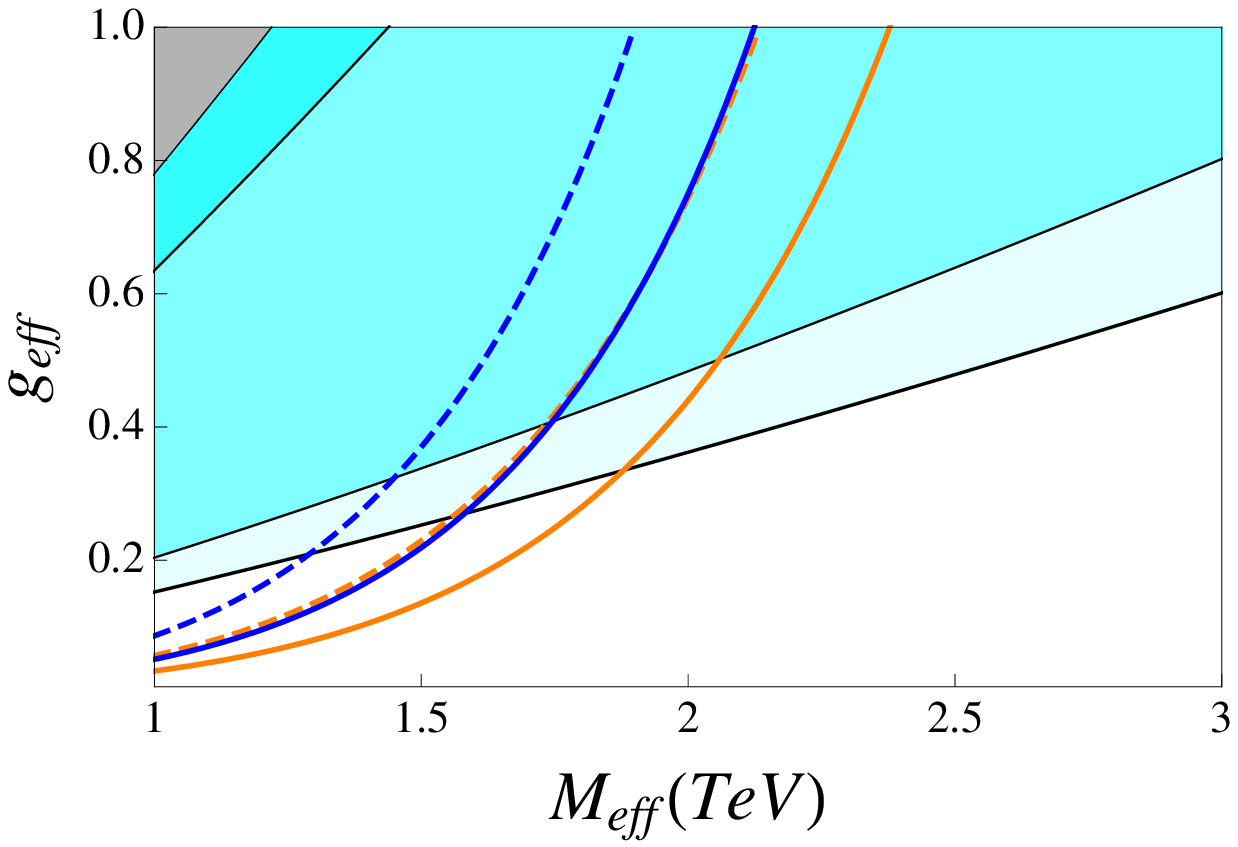}
\end{minipage}
\vspace{0.2 cm}
\caption{As fig. \ref{fig:Lim_Sim}, but for production of two
  leptoquark scalars $S_{2/3}^{LQ}$ (blue) and $S_{1/3}^{LQ}$
  (orange). Note, that the dashed line for $Br(S \to e e j j ) =
  10^{-2}$ in case of $S_{1/3}^{LQ}$ is very similar to $Br(S \to e e
  j j ) = 10^{-1}$ for the case of $S_{2/3}^{LQ}$. }
 \label{fig:Lim_LQ}
\end{figure}

In Fig. \ref{fig:Lim_Sim}, \ref{fig:Lim_LQ} we then plot the
sensitivities of $\znbb$ decay and the LHC for five different cases in
the plane $g_{eff}$ versus $M_{eff}$.  For the LHC we show the
expected sensitivity limits, assuming less then 3 signal events in 300
fb$^{-1}$ of statistics, and plot for two values of ${\rm Br}(S\to
eejj)$, i.e $10^{-2}$ (dashed lines) and $10^{-1}$ (solid lines) ,
for two different values of $m_{\psi}=200$ GeV (left) and $m_{\psi}=1$
TeV (right).  Again, for larger masses $m_{\psi} $ the LHC limits
become more stringent except for the region $(m_{\psi} - m_{S})
\lesssim 100$ GeV, where the LHC sensitivity is low. The
different color codes correspond to the five different scalar bosons,
that can be singly produced at the LHC, namely $S_{+1}$ (red),
$S^{DQ}_{4/3} $ (black), $S^{DQ}_{2/3}$ (purple), $S^{LQ}_{2/3}$
(blue) and $S^{LQ}_{1/3}$ (orange).  In Fig. \ref{fig:Lim_Sim} we have
ploted three cases which correspond to the scalars $S_{+1}$,
$S^{DQ}_{4/3}$ and $S^{DQ}_{2/3}$ while in Fig. \ref{fig:Lim_LQ} we
have plotted the remaining two leptoquark cases $S^{LQ}_{2/3}$ and
$S^{LQ}_{1/3}$.  In addition Figs. \ref{fig:Lim_Sim}, \ref{fig:Lim_LQ}
show four different cases for current and future limits from $\znbb$
decay. The dark gray area is, as in fig \ref{fig:Lim_8TeV}, the
currently excluded part of parameter space from non-observation of
$^{136}$Xe decay with $T_{1/2}^{\znbb}\ge 1.6 \times 10^{25}$ ys
\cite{Auger:2012ar} assuming $\znbb$ decay is caused by the
decomposition with the smallest rate (see Table \ref{Tab:bbstat}), and
thus corresponds to the most pessimistic case for the sensitivity of
$\znbb$ decay. The three blue areas are (from left to right): Smallest
rate, but for a limit of $T_{1/2}^{\znbb}\ge 10^{26}$ ys, largest rate
with $T_{1/2}^{\znbb}\ge 10^{26}$ ys and, finally the largest rate
with $T_{1/2}^{\znbb}\ge 10^{27}$ ys. The lightest area to the right
therefore corresponds to the most optimistic reach for $\znbb$ decay
in the foreseeable future.

As can be seen from Figs. \ref{fig:Lim_Sim}, \ref{fig:Lim_LQ}, with
the exception of the LQ cases (Fig.  \ref{fig:Lim_LQ}), the LHC at
$\sqrt{s}=14$ TeV will be more sensitive than $\znbb$ decay
experiments as probe for LNV. For the LQ case, the LHC is more
sensitive than $\znbb$ decay in the pessimistic case for $\znbb$
(operators ${\cal O}_1$ and ${\cal O}_5$ in the notation of
\cite{Pas:2000vn}) but not for the one to which $\znbb$ decay is most
sensitive, particularly ${\cal O}_2$.  For the remaining operators
${\cal O}_3$ and ${\cal O}_4$ $\znbb$ decay and LHC sensitivities are
very similar.

Now we will discuss the second group of decompositions, which
correspond to the "asymmetric'' cases, with two different scalar
masses and all couplings different.  In this case, the assumption
$g_1=g_4$, $g_2=g_3$ and $m_{S}= m_{S^\prime}$ in
Eq. (\ref{Efective-S1}) is violated and the plane $g_1^2$ vs $m_S$ is
more adequate for comparison of LHC and $\znbb$ decay sensitivities.

\begin{figure}[htbp]
\begin{minipage}[b]{.45\linewidth}
\includegraphics[width=\linewidth]{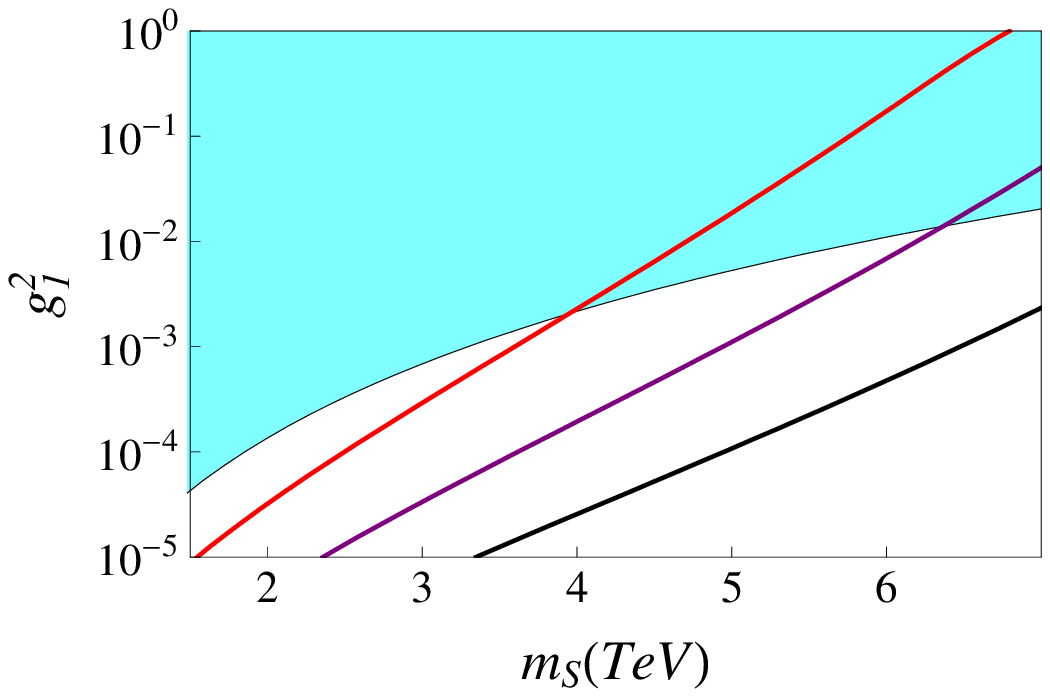}
\end{minipage}
\begin{minipage}[b]{.05\linewidth}
\hspace{1pt}
\end{minipage}
\begin{minipage}[b]{.45\linewidth}
\vspace{0pt}
\includegraphics[width=\linewidth]{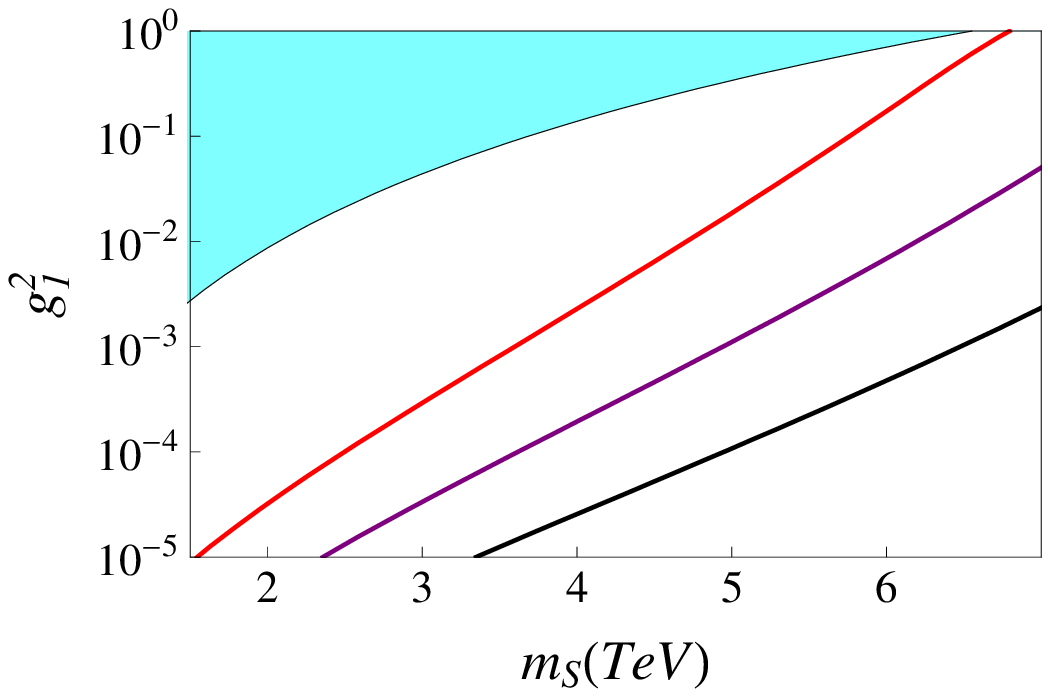}
\end{minipage}
\vspace{0.2 cm}
\caption{ Future limits for the LHC at $\sqrt {s} = 14$ TeV compared
  with future double beta decay experiments.  The blue region
  represents the parameter region accessible in near future $\znbb$
  experiments, whereas the colored lines shows sensitivity limits for
  the LHC for production of three different scalar bosons $S_{+1}$
  (red), $S_{2/3}^{DQ}$ (purple) and $S_{4/3}^{DQ}$ (black). Solid
  lines were calculated using $Br(S \to e e j j ) = 10^{-1}$ and the
  blue region was calculated using $m_{\psi} = 1.5$  TeV,
  $m_{S^\prime} = 2.0$ TeV, $g_2 = g_3 = g_4 = 1$ (left) and $g_2 =
  g_3 = g_4 = 0.5$ (right).  }
 \label{fig:Lim_Asim}
\end{figure}

In Fig. \ref{fig:Lim_Asim} we then compare the sensitivities of
$\znbb$ decay and the LHC for three different cases, using
Eq. (\ref{eq:SigBrS1}). The different color codes correspond to the
three different scalar bosons, that can be singly produced at the LHC,
namely $S_{+1}$ (red), $S^{DQ}_{4/3} $ (black), $S^{DQ}_{2/3}$
(purple).  For the LHC we show the expected sensitivity limits for
$Br(S \to e e j j ) = 10^{-1}$ (solid lines) in the plane $g_{1}$
versus $m_{S}$.

Fig. \ref{fig:Lim_Asim} shows future limits from $\znbb$ decay which
corresponds to the most optimistic reach for $\znbb$ decay in the
foreseeable future. Those limits were calculated using, in Eq.
(\ref{eq:meff}), $m_{\psi} = 1.5$ TeV , $m_{S^\prime} = 2.0$ TeV, $g_2
= g_3 = g_4 = 1$ (left) and $g_2 = g_3 = g_4 = 0.5$ (right). For
larger masses $m_{\psi}, m_{S^\prime}$ or smaller couplings $g_2, g_3,
g_4 $ those limits become weaker. The choice of $m_{\psi} = 1.5$ TeV,
$m_{S^\prime} = 2.0$ TeV is reasonable since all the "asymmetric"
decompositions must have coloured fermions and these can be
constrained through pair production searches, which will yield
sensitivity limits on their masses of $2-2.5$ \ TeV.  Moreover in the
"asymmetric" decompositions the scalar $S^{\prime}$ is a leptoquark or
a $S_{+2}$. Also for leptoquarks the LHC searches from pair and single
productions \cite{CiezaMontalvo:1998sk} will have sensitivities around $2$ TeV
and the doubly charged scalar $S_{+2}$ can also be searched trough pair
production (through a production graph with a virtual photon). As can
be seen from Fig. \ref{fig:Lim_Asim} the LHC at $\sqrt{s}=14$ TeV will
be more sensitive than $\znbb$ decay experiments as probe for LNV for
all the "asymmetric" decompositions.

\begin{figure}[h]
  \includegraphics[scale=0.8]{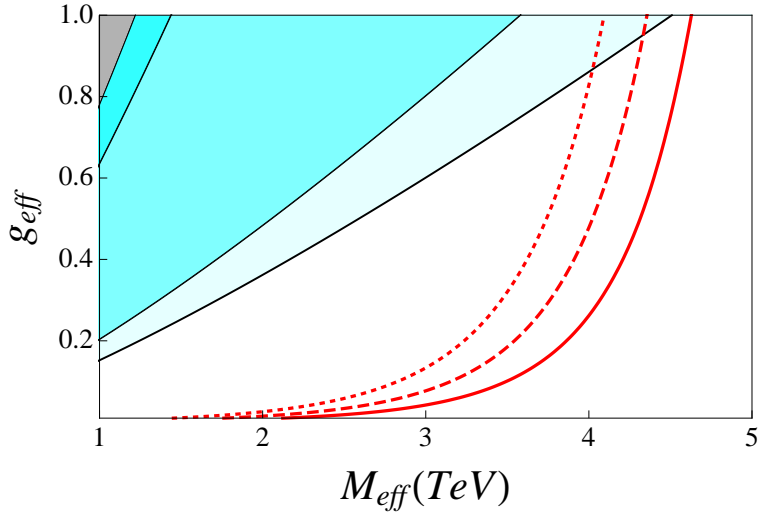}
  \caption{Comparison of expected sensitivity limits assuming less
    than 3 (solid), 10 (dashed) and 30 (dotted) signal events in $300
    fb^{-1}$ of statistic at LHC for the production of the scalar
    bosons $S_{+1}$.  Red lines were calculated using $Br(S \to
    e e j j ) = 10^{-1}$ and fermion mass $m_{\psi_0} = 1000$
    GeV. The gray region on the top left corner is ruled out by
    $\znbb$ whereas the blue region represents the parametric region
    accessible in near future $\znbb$ experiments. See text for more
    details.}
 \label{fig:Sens}
\end{figure}

Finally we have compared in Fig. \ref{fig:Sens} sensitivity limits for
the "symmetric" decomposition $(\bar u d)(\bar e )(\bar e)(\bar u d)$
assuming 3, 10 and 30 events in $300 fb^{-1}$ of statistics. As one can
see from Fig. \ref{fig:Sens} even under the pessimistic 
assumption that only 30 signal events can be excluded, our previous 
limits calculated for the more optimistic situation of 3 events 
(see Fig. \ref{fig:Lim_Sim}) suffer only minor changes (in this 
linear plot) and the LHC is still more sensitive than $\znbb$.
More accurate numbers of the total number of events necessary to 
claim discovery/exclusion would require a full detector MonteCarlo, 
outside the scope of this paper.

\section{Distinguishing LNV models at the LHC}
\label{Sec:Dst}

In the previous section we have compared the sensitivity of the LHC
with $\znbb$ decay. Here, we discuss the question how the different
LNV decompositions could actually be distinguished using LHC data, if
a positive signal were to be found in the $\sqrt{s}=14$ TeV run. We
will consider two types of observables: (i) charge asymmetry 
\footnote{This charge asymmetry has also been discussed in a 
different context in the recent paper \cite{Durieux:2012gj}.}  and
(ii) invariant mass peaks. Interestingly, the combination of the two
sets of observables is sufficient to distinguish among nearly all
decompositions. The only exceptions are the pairs of cases
(1-ii-a)-(1-ii-b) and (1-i)-(3-i), the latter, however, only in the
``mass-degenerate'' limit, see below.

Recall first that the scalars $S_{+1}$, $S^{DQ}_{4/3}$ and
$S^{DQ}_{2/3}$ are produced in s-channel, while single leptoquarks are
produced at the LHC always in association with a lepton. The LQ final
state that we are interested in, is therefore $eejjj$, different from
the other cases, see discussion next section.  We will therefore
separate the discussion here in ``LQ-like'' and other cases.

\subsection{Charge asymmetry}
\label{SubSec:CA}

In the dilepton event samples, there are three subsets of events with 
different charges: $e^+e^+$, $e^+e^-$ and $e^-e^-$. From these three 
numbers we can form two independent ratios:
\begin{eqnarray}\label{eq:ca}
x_{CA} = \#(e^+e^+)/\#(e^-e^-) \\ \nonumber
y_{CA} = \#(e^-e^+)/\#(e^+e^+) .
\end{eqnarray}
Consider the simpler case of $y_{CA}$ first. In the cases where the
fermion in the diagram is neutral, $\psi=\psi_0$, it is a Majorana
particle and at tree-level ${\rm Br}(\psi_0\to e^+ jj)={\rm
Br}(\psi_0\to e^- jj)$.
\footnote{The branching ratios equal $1/2$ in case there is no 
generation mixing.} Thus, all decompositions with $\psi_0$ will 
have $y_{CA}=1/2$, up to loop corrections. The situation is different 
for decompositions with charged fermions. Here we can distinguish the 
cases involving $\psi_{4/3}$ and $\psi_{5/3}$, on the one hand, and 
$\psi_{1/3}$ on the other hand. Since $\psi_{4/3}$ and $\psi_{5/3}$ 
can decay only into $e^+e^+j$ and $e^-e^-j$, all decompositions involving 
these fermions have $y_{CA}=0$. Finally, $\psi_{1/3}$ can decay into 
both charge signs, but the branching ratio of $\psi_{+1/3}$ into positrons 
and electrons involve different combinations of couplings (and masses), 
and therefore are free numbers. $y_{CA}$ in this case is arbitrary, 
but could be used to fix some combination of couplings experimentally.

We now turn to the discussion of $x_{CA}$. Define the ratio for 
the LHC production cross section of one of our five scalars, relative 
to the cross section for its charge conjugate state as:
\begin{equation}\label{eq:Rsig}
R_{\sigma}^{S_i} = \frac{\sigma(pp\to S_i)}{\sigma(pp\to {\bar S_i})},
\end{equation}
Here, $S_i$ stands for any of $S_i=S_{+1},S^{DQ}_{4/3},S^{DQ}_{2/3},
S^{LQ}_{2/3},S^{LQ}_{1/3}$.  We can divide the discussion of all 18
decompositions into three groups of cases. We put into the first 
group the six decompositions without any leptoquark in the diagram, 
i.e. all decompositions T-I-i and T-I-iii of table \ref{Tab:TopoI}.
Into the second group we put the decompositions with two leptoquarks, 
i.e. 2-iii, 4-i and 5-i. The remaining 8 decompositions with one 
leptoquark form the third group.
\begin{figure}[htb]
\centering
\includegraphics[width=0.5\linewidth,height=0.5\linewidth]{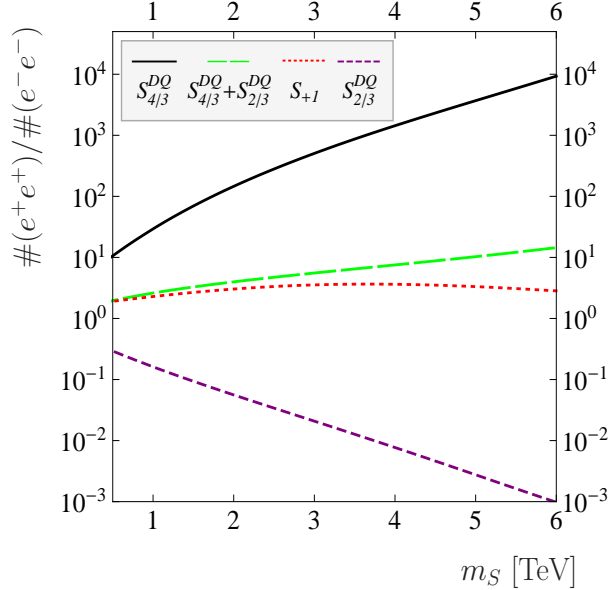}
\vskip-3mm
\caption{\label{fig:CASmplst}Charge asymmetry $x_{CA}$, see eq. (\ref{eq:ca}), 
as a function of the boson mass for different kinds of scalars. 
Shown are the cases with $S_{+1}$ or diquarks, for discussion see text.}
\end{figure}  

We start the discussion with group-(1). As shown in
fig. (\ref{fig:xsectTI}), $R_{\sigma}^{S_i}$ is different for the
various scalars and moreover strongly dependent on the mass of the
scalar. \footnote{Note that $R_{\sigma}^{V_i}$ for vectors will behave
exactly as $R_{\sigma}^{S_i}$ discussed here.} This asymmetry in cross
sections will cause the charge asymmetry $x_{CA}$ to depend strongly
on the decomposition.  The charge asymmetry $x_{CA}$ is shown in
fig. \ref{fig:CASmplst} for diquarks and for $S_{+1}$.  Consider first
the case denoted $S_{+1}$ on the left. This corresponds to both, the
case 1-i and the two sub-cases 1-ii-a and 1-ii-b. The former is an
example of a symmetric decomposition, i.e. here two of the four
couplings in the diagram are pairwise equal, namely $g_{S_{+1}{\bar
u}d}$ connecting two outer legs and and $g_{S_{+1}{\bar e}\psi_0}$
connecting two propagators in either beta decay subprocess to the left
and to the right. It is straightforward to show that upon calculating
$x_{CA}$ all couplings cancel out an $x_{CA}$ simply is $x_{CA}^{\rm
sym} = R_{\sigma}$. Thus, for symmetric decompositions, $x_{CA}$ at
any fixed mass of the scalar is simply a number, predicted by the
decomposition. \footnote{We note again that the classical case of LR
symmetry has the same $x_{CA}$ as shown here for $S_{+1}$.} The two
sub-cases 1-ii-a and 1-ii-b can be called ``absolutely asymmetric''
decompositions, since the $S_{+2}$ can not be singly produced in the
LHC. In these cases only the couplings at the $S_{+1}$ vertices matter
and drop out again in the calculation of $x_{CA}$. Thus, the line
denoted $S_{+1}$ in fig. (\ref{fig:CASmplst}) is valid for both,
decomposition 1-i and 1-ii.

Consider DQs, i.e. decompositions 3-i, 3-ii and 3-iii. The lines
denoted $S^{DQ}_{4/3}$ and $S^{DQ}_{2/3}$ in fig. (\ref{fig:CASmplst})
correspond to decompositions 3-ii and 3-iii. In these two cases the
decomposition is ``completely asymmetric'' and therefore does not
depend on the values of individual couplings, but depends strongly on
the mass of the scalar.

For decomposition 3-i the discussion is slightly more
complicated. Here, one can distinguish the case where the two diquark
masses are degenerate ($m_{S^{DQ}_{4/3}}=m_{S^{DQ}_{2/3}}$) and the
non-degenerate case.  In the non-degenerate case the distributions in
$m_{eejj}^2$ would show two distinct peaks, one having the $x_{CA}$
appropriate for $S^{DQ}_{4/3}$ while the other has $x_{CA}$ of
$S^{DQ}_{2/3}$. In fact, such a non-degenerate case is not only
``easy'' to resolve from the decomposition(s) involving $S_{+1}$,
having more than one peak in $m_{eejj}^2$ would actually allow to
probe for all four couplings entering the diagram and thus provide
more information than in other cases. In the mass degenerate limit,
however, both $S^{DQ}_{4/3}$ and $S^{DQ}_{2/3}$ contribute to the
number of events in the same peak. In this case, $x_{CA}$ depends on
the relative ratio of coupling of the two diquarks to
fermions. Fig. (\ref{fig:CASmplst}) shows $x_{CA}$ for this case,
$S^{DQ}_{4/3}+S^{DQ}_{4/3}$ in the limit where the diquark couplings
to fermions are equal. For arbitrary ratios of couplings (but
degenerate masses) $x_{CA}$ can vary between the two extreme limits
shown as $S^{DQ}_{4/3}$ and $S^{DQ}_{2/3}$. Measurement of $x_{CA}$
anywhere between those two extremes, therefore points toward
decomposition 3-i in case of DQs. The problematic case for
distinguishing between 3-i and decomposition 1-i is therefore the mass
degenerate case for decomposition 3-i, where the two pairs of diquark
couplings conspire to give a $x_{CA}$ equal (or very similar) to the
corresponding one for $S_{+1}$.

\begin{figure}[htb]
\centering
\includegraphics[width=0.45\linewidth,height=0.5\linewidth]{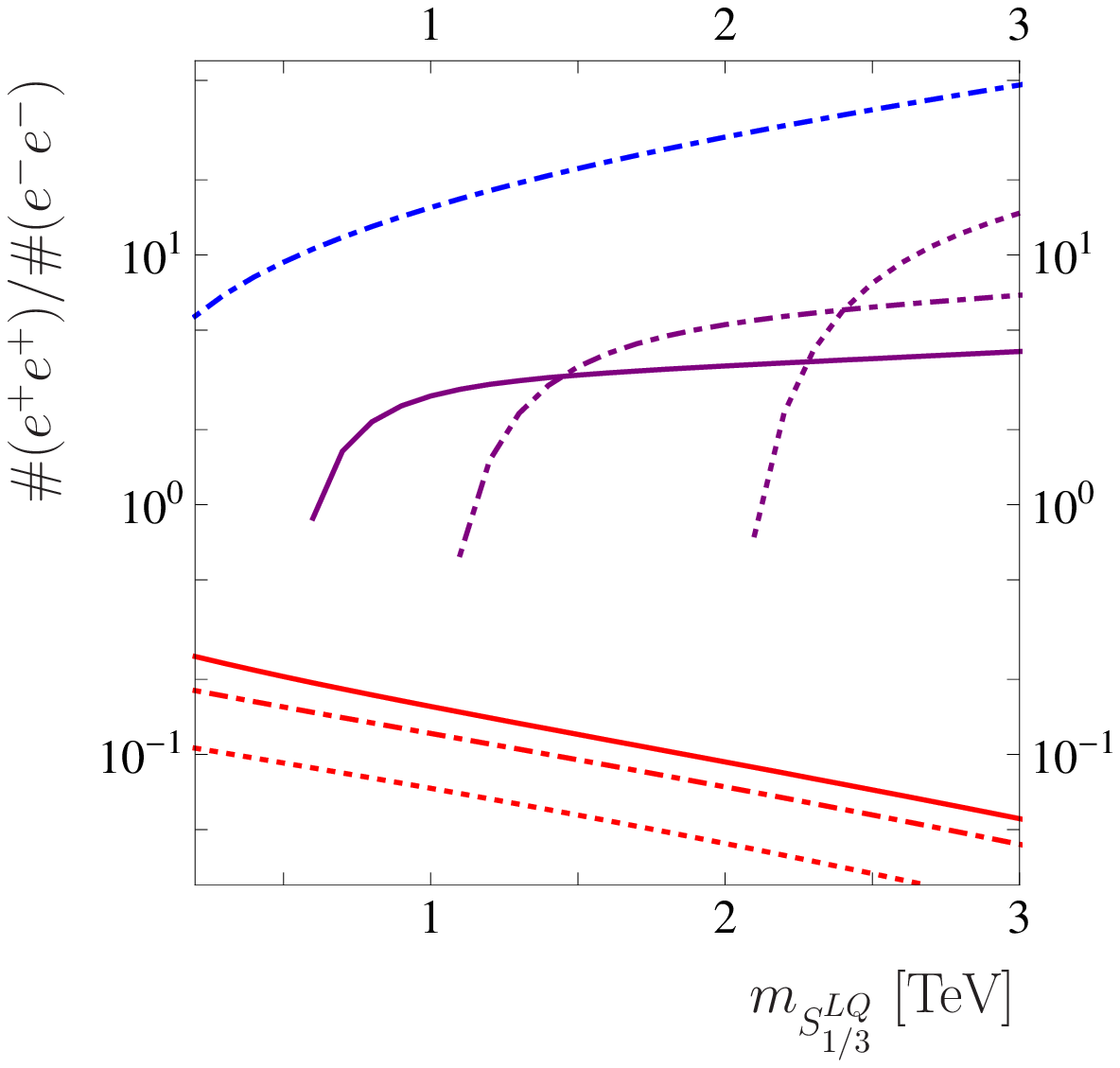}
\includegraphics[width=0.45\linewidth,height=0.5\linewidth]{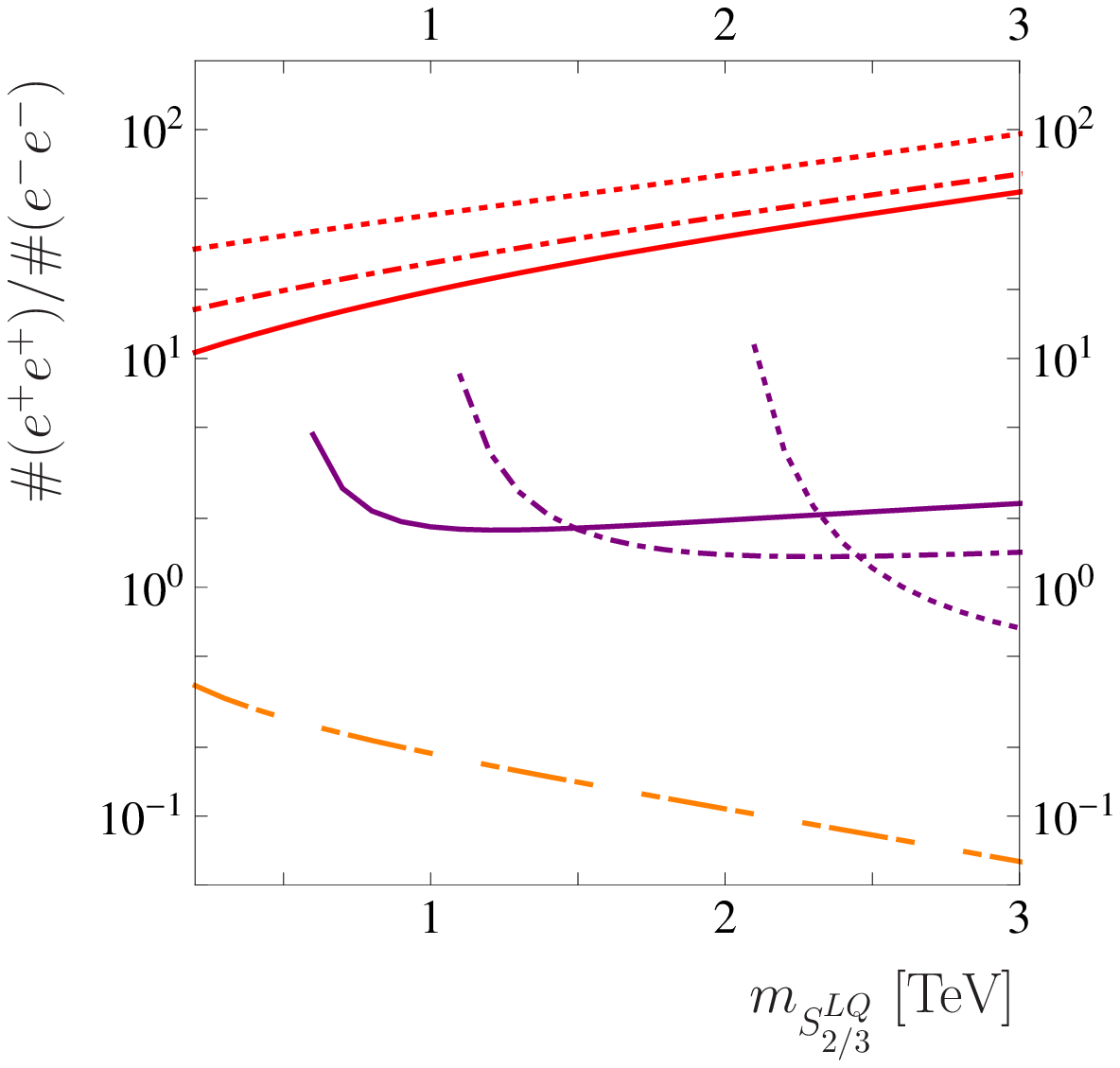}
\vskip-3mm
\caption{\label{fig:CALQ}  Charge asymmetry $x_{CA}$ as a function of 
the leptoquark mass, to the left $S^{LQ}_{1/3}$, to the right $S^{LQ}_{2/3}$. 
The different lines show different cases: blue (left, dash-dotted) and 
orange (right, dash-dotted) show $x_{CA}$ for $S^{LQ}_{1/3}+e$ and 
$S^{LQ}_{2/3}+e$ production, respectively. The red lines in both plots 
show  $x_{CA}$ for $S^{LQ}_{q}+\psi$ production only. The purple lines 
show in both cases $x_{CA}$ combining both production modes, assuming 
the couplings are equal, $g_{eQ S^{LQ}_{q}}=g_{\psi Q' S^{LQ}_{q}}$. 
Shown are three calculations for different values of $m_{\psi}$: 
$m_{\psi}=0.5$ TeV - full lines; $m_{\psi}=1$ TeV - dot-dashed lines 
and   $m_{\psi}=2$ TeV - dotted lines. In all cases the calculation 
includes a phase space suppression for Br($S^{LQ}_{Q}\to q+\psi$) as 
described in the text.}
\end{figure}  

We now turn to the discussion of $x_{CA}$ for the four decompositions 
with two LQs. As mentioned previously, the final states for LQs 
are $e^-e^-jjj$ and $e^+e^+jjj$. LNV with LQs can therefore, 
in principle, be distinguished from DQs and $S_{+1}$. However, 
in the discussion of $x_{CA}$ for LQs one more complication arises: 
The final LNV states can be produced via two different intermediate 
states, i.e. $S^{LQ}_{q}+e$ and $S^{LQ}_{q}+\psi$. In case of 
$S^{LQ}_{2/3}$, for example, the main production diagram is 
$d+g \to S^{LQ}_{q}+e^{-}$, contributing to $e^-e^-$, while 
$u+g\to S^{LQ}_{2/3}+\psi$ will contribute to $e^+e^+$-like 
events. For the case of $\sigma(pp\to S^{LQ}_{q}+\psi)$, the 
cross section not only depends strongly on $m_{S^{LQ}_{q}}$, 
but also depends on $m_{\psi}$, see fig. (\ref{fig:xsectTILQ}). 
However, also in the case of $S^{LQ}_{q}+e$ production, the 
mass of $\psi$ enters in the calculation of the total number of 
events, since the branching ratio of Br($S^{LQ}_{q}\to \psi +z$), 
where $z$ stands for all possible SM fermion states, suffers a 
phase space suppression factor for large $m_{\psi}$:
\begin{equation}\label{eq:phsp}
f(m_S^2,m_{\psi}^2) = \frac{(m_S^2-m_{\psi}^2)^2}{m_S^4}.
\end{equation}
The predicted charge asymmetry then depends on whether events 
from $\sigma(pp\to S^{LQ}_{q}+\psi)$ can be separated from 
$\sigma(pp\to S^{LQ}_{q}+e)$ or not. This separation can be 
done, in principle, by looking at the invariant mass peaks 
discussed in the next section. However, especially in case the 
total number of events is low, such a separation will become 
difficult (and inefficient). Then the charge asymmmetry 
measured will be an averaged charged asymmetry of both 
production modes. 

In fig. (\ref{fig:CALQ}) we plot calculated $x_{CA}$ as a function of
the leptoquark mass, to the left $S^{LQ}_{1/3}$, to the right
$S^{LQ}_{2/3}$ for a number of cases. The blue (left, dash-dotted) and
orange (right, dash-dotted) lines show $x_{CA}$ for $S^{LQ}_{1/3}+e$
and $S^{LQ}_{2/3}+e$ production, respectively. The red lines in both
plots show $x_{CA}$ for $S^{LQ}_{q}+\psi$ production only.  Shown are
three calculations for different values of $m_{\psi}$: $m_{\psi}=0.5$
TeV - full lines; $m_{\psi}=1$ TeV - dot-dashed lines and $m_{\psi}=2$
TeV - dotted lines. These lines are the predicted $x_{CA}$ for the
case that events from $S^{LQ}_{q}+\psi$ can be separated completely
from those stemming from $S^{LQ}_{1/3}+e$. In the more conservative
case that averaging over both production modes has to be done, the
predicted $x_{CA}$'s are plotted as purple lines, again for three
different values of $m_{\psi}$. In this calculation we assumed for
simplicity that $g_{eQ S^{LQ}_{q}}=g_{\psi Q' S^{LQ}_{q}}$ and
included a phase space suppression factor in the calculation of events
for $S^{LQ}_{1/3}+e$ production to account for the reduced
Br($S^{LQ}_{q}\to \psi +z$). This phase space suppression, see
eq. (\ref{eq:phsp}), is responsible for the sharp bend in the lines at
low $m_{S^{LQ}_q}$. For larger or smaller ratios of $g_{eQ
S^{LQ}_{q}}$ to $g_{\psi Q' S^{LQ}_{q}}$ the corresponding 
lines for $x_{CA}$ will change, thus $x_{CA}$ can vary in principle 
between the extremes shown in fig. (\ref{fig:CALQ}) for arbitrary 
values of the couplings. However, if $m_{S^{LQ}_q}$ and $m_{\psi}$ 
are known from measurement, the ratio of $g_{\psi Q' S^{LQ}_{q}}$ 
and $g_{eQ S^{LQ}_{q}}$ can in principle be fixed from a measurement 
of Br($S^{LQ}_{q}\to \psi +z$).

The above discussed results cover the decompositions 4-i and 5-i, in
which the two LQs in the diagram have the same quantum numbers.  For
the case of decompositions 2-iii, on the other hand, both types of LQs
contribute and the resulting $x_{CA}$ will be the average of the
individual $x_{CA}$'s shown in fig. (\ref{fig:CALQ}) on the left and
right. For all couplings equal, the resulting $x_{CA}$ varies smoothly 
from around $x_{CA}=2$ for $m_{S^{LQ}_q} = 1$ TeV to $x_{CA} \simeq 3$ 
for $m_{S^{LQ}_q} = 3$ TeV, for decomposition 2-iii-a. 
There is, however, one subtle difference between decomposition
2-iii-a and 2-iii-b, since in these two the up and down quarks in the
initial state for the case of $\sigma(pp\to S^{LQ}_{q}+\psi)$ 
production is interchanged. This leads to slightly lower values for 
$x_{CA}$ in 2-iii-b compared to 2-iii-a.

Finally, we briefly discuss the remaining eight decompositions within
group-(3). In this case, in principle, mass peaks should show up in
$m_{eejj}^2$ {\em and} $m_{e_2jjj}^2$, clearly identifying the LQ and
the other scalar boson by the individual $x_{CA}$'s. However, this
statement assumes that cross sections for LQs are large enough that
for these decompositions both types of scalars are produced at the
LHC. Considering the large ratio of cross sections for $S_{+1}$ and
DQs relative to cross sections for LQs, this might be a too optimistic
assumption. Thus, in the case only one peak in $m_{eejj}^2$ is found,
only the ``leading'' boson of the decomposition can be identified and
there appears a degeneracy among the decompositions in group-(1) and
group-(3) in this observable.

\subsection{Invariant mass peaks}
\label{SubSec:MP}

We now turn to the discussion of differentiating between different 
decompositions using peaks in the cross sections in different experimentally 
measurable invariant mass systems. Again, we will divide this discussion 
into different cases. First, we will discuss decompositions with at 
least one $S_{+1}$, then decompositions with at least one diquark. 
These two cases can be distinguished, in principle, by measuring the 
charge asymmetry discussed in the last subsection. Finally, we will 
discuss decompositions which contain only LQs.

\begin{table}[h!]  
\begin{center}
\begin{tabular}{|c|c|c|c|c|c||c|c|c|c|}\hline
Case & $m_{S}$ & $m_{\psi}$ & $m_{S'}$ & Decomposition\\
\hline
A & $m(eejj)$ & $m(ejj)$ & $m(jj)$ & $(\bar u d)(\bar e)(\bar e)(\bar u d)$\\
\hline
B & $m(eejj)$ & $m(ejj)$ & $m(ej)$ & 
  $({\bar u}d)({\bar e})({\bar u})(d{\bar e})$ ; 
  $( \bar u d ) ( \bar e  ) ( d  ) ( \bar u \bar e  )$  \\
\hline
C & $m(eejj)$ & $m(eej)$ & $m(ee)$   & 
  $(\bar u d   ) ( \bar u  ) ( d ) ( \bar e \bar e )$ ; 
  $( \bar u d ) ( d ) (\bar u  ) ( \bar e \bar e  )$ \\
\hline
D & $m(eejj)$ & $m(eej)$ & $m(ej)$  & 
  $(\bar u d   ) ( \bar u  ) ( \bar e ) ( d  \bar e )$ ; 
  $( \bar u d ) ( d ) (\bar e  ) ( \bar u \bar e  )$ \\
\hline
\end{tabular}
\end{center}
\caption{\it Combinations of invariant mass distributions where peaks
in the cross sections arise, in case the mass ordering is
$m_S>m_{\psi}>m_{S'}$, for decompositions of $\znbb$ decay with charge
asymmetries that are " $(\bar u d)$ like". If $m_{\psi}\le m_{S'}$, 
$m_{S'}$ can not be measured and cases A=B and C=D can not be distinguished 
by this observable.
\label{InvMassUD}}
\end{table}

Table \ref{InvMassUD} shows the results of the analysis for
decompositions involving $S_{+1}$. With the exception of case (A),
where $S_{+1}$ appears twice in the diagram, the two scalars in the
decomposition are different particles. In case $m_S>m_{\psi}>m_{S'}$,
there are then two subsystems in the sample of $eejj$-events which
form mass peaks and we can distinguish cases (A)-(D), leaving only
``degeneracies'' in decompositions (1-ii-a)-(1-ii-b) (corresponding to
case B) and (2-iii-a)-(2-iii-b) (case C), where both mass peaks and
$x_{CA}$ are (pairwise) identical.  However, it is possible that
$m_{\psi}\le m_{S'}$, in which case $m_{S'}$ can not be measured and
cases A=B and C=D can not be distinguished anymore.

\bigskip

\begin{table}[h!]  
\begin{center}
\begin{tabular}{|c|c|c|c|c|c||c|c|c|c|}\hline
Case & $m_{S}$ & $m_{\psi}$ & $m_{S'}$ & Decomposition\\
\hline
A & $m(eejj)$ & $m(ejj)$ & $m(jj)$  & 
  $(\bar u \bar u) (\bar e)(\bar e) (d d)$\\
\hline
B & $m(eejj)$ & $m(ejj)$ & $m(ej)$  &  
  $(\bar u \bar u   ) ( \bar e  ) ( d ) ( d \bar e )$ ; 
  $( d d ) ( \bar e ) (\bar u  ) ( \bar u \bar e  )$\\
\hline
C & $m(eejj)$ & $m(eej)$ & $m(ee)$ & 
   $(\bar u \bar u   ) ( d  ) ( d ) ( \bar e \bar e )$ ; 
   $( d d ) ( \bar u ) (\bar u  ) ( \bar e \bar e  )$\\
\hline
D & $m(eejj)$ & $m(eej)$ & $m(ej)$  &   
  $(\bar u \bar u   ) ( d  ) ( \bar e ) ( d \bar e )$ ; 
  $( d d ) ( \bar u ) (\bar e  ) ( \bar u \bar e  )$\\
\hline
\end{tabular}
\end{center}
\caption{\it As above, for decompositions of $\znbb$ decay with charge
  asymmetries that are "$(\bar u \bar u)$ and $(d d) $ like". If
  $m_{\psi}\le m_{S'}$, $m_{S'}$ cannot be measured and cases A=B and
  C=D can not be distinguished with this observable.
\label{InvMassUU}}
\end{table}

Table \ref{InvMassUU} shows the results of the analysis for
decompositions involving DQs. Case (A) has the same invariant mass 
peaks as case (A) in table  \ref{InvMassUD}. Thus, in case 
$m_{S^{DQ}_{4/3}}=m_{S^{DQ}_{2/3}}$ the decomposition (3-i) can 
not be distinguished from (1-i), if also the diquark couplings ``conspire'' 
such that $x_{CA}$ agrees with the corresponding value for 
$S_{+1}$. In this case, the only difference between (3-i) and 
(1-i) is that (3-i) always requires a electrically charged 
coloured fermion, which could show up in pair production.
Cases (B)-(D) in table  \ref{InvMassUU} are also equal to 
(B)-(D) in table  \ref{InvMassUD}. However, in these cases 
DQ decompositions and $S_{+1}$ decompositions can always 
be distinguished by measuring $x_{CA}$.

\begin{table}[h!] 
\label{InvMass} 
\begin{center}
\begin{tabular}{|c|c|c|c|c|c||c|c|c|c|}\hline
Case & $m_{S}$ & $m_{\psi}$ & $m_{S'}$ & Decomposition\\
\hline
A & $m(e_2jjj)$ & $m(e_2jj)$ & $m(jj)$  &  
  $(\bar u \bar e) ( d)(\bar e) (\bar u d)$ ; 
  $(\bar u \bar e) (\bar u)(\bar e) (d d)$  ; 
  $(d  \bar e) (\bar u)(\bar e) (\bar u d)$ ; 
  $(d \bar e) (d)(\bar e) (\bar u \bar u)$\\
\hline
B & $m(e_2jjj)$ & $m(e_2jj)$ & $m(e_2j)$  & 
  $(\bar u \bar e) ( d)(\bar u) ( d \bar e )$ ; 
  $(\bar u \bar e) (d )( d) ( \bar u \bar e)$  ; 
  $(d  \bar e) (\bar u)(\bar u) ( d \bar e)$ ; 
  $(d \bar e) (d)(\bar u) ( \bar u \bar e)$\\
\hline
C & $m(e_2jjj)$ & $m(jjj)$ & $m(jj)$  & 
  $(\bar u \bar e) ( \bar e)(\bar u) ( d d )$ ; 
  $(\bar u \bar e) ( \bar e )( d) ( \bar u d)$ ; 
  $(d  \bar e) (\bar e)(\bar u) (  \bar u d)$ ; 
  $(d \bar e) ( \bar e)(\bar u) ( \bar u d)$\\
\hline
\end{tabular}
\end{center}
\caption{\it As above, for decompositions of $\znbb$ decay with charge
  asymmetries that are "$(\bar u \bar e)$ and $(d \bar e)$ like",
  i.e. for single leptoquark production. Recall that the the complete
  signal is ``$eejjj$'' without a peak in $m_{eejjj}^2$. If
  $m_{\psi}\le m_{S'}$, cases A=B cannot be distinguished. Note that
  case B, where both sides of the decomposition contain only
  leptoquarks, will produce ``$eejjj$'' final states, only. In all
  other cases, also the ``$eejj$' signal should arise.
\label{InvMassLQ}}
\end{table}

Finally, table \ref{InvMassLQ} shows combinations of invariant mass
distributions in case the mass ordering is $m_S>m_{\psi}>m_{S'}$, for
decompositions of $\znbb$ decay with charge asymmetries that are
"$(\bar u \bar e)$ and $(d \bar e)$ like", i.e. for single leptoquark
production.  Here, it is assumed that events from $S^{LQ}_{q}+e$
production can be distingsuished from $S^{LQ}_{q}+\psi$
production. The table refers to the former. (Note again, if
$m_{\psi}\le m_{S'}$, cases A=B can not be distinguished, leaving a
degeneracy in the identification of the decomposition in that case.)
In case of events from $S^{LQ}_{q}+\psi$ production, the decay of the
LQ will lead to a peak in $m_{e_aj_x}^2=m_{S^{LQ}_{q}}^2$ and the
decay of $\psi$ to $m_{e_bj_yj_z}^2=m_{\psi}^2$. Note, however, that
in case of $S^{LQ}_{q}+e$ production, the mass peak for $m_{\psi}^2$
is formed by a subsystem of ``$e_2jjj$'' (which gives
$m_{S^{LQ}_{q}}^2$), whereas in $S^{LQ}_{q}+\psi$ production the two
mass peaks must come from different leptons and jets, i.e.  $a \ne b$
and $y \ne x \ne z$. This feature can be used to separate
$S^{LQ}_{q}+e$ from $S^{LQ}_{q}+\psi$ production.

Before closing, we briefly mention pair production of coloured
fermions.  Here, a signal $eejjjj$, i.e. at least four hard jets,
would test the different decompositions of double beta decay. We note
that decompositions with $\psi_0$ exist with a colour singlet fermion,
for which pair production at the LHC is negligible, while for all the
12 ``new'' decompositions, see discussion in section \ref{Sec:Dec} and
section \ref{sect:xsect}, pair production of the exotic fermions is
expected to probe the existence of such states with masses up to
roughly $m_{\psi} \sim 2-2.5$ TeV, depending on the final state
branching ratios. Note, that also pair produced $\psi$ can, depending
on decomposition, produce in some cases like-sign dileptons. Also, if
a signal is found in $eejjjj$, there is a threshold for these events 
at $m_{eejjjj}=2 m_{\psi}$. Different subsystems again form mass peaks 
at $m_{\psi}$ and in case, this fermion is heavier than (one of)
the bosons mass peaks in sub-sub-systems will show up, providing
additional information. The possible combinations can be straightforwardly 
derived from table \ref{Tab:TopoI}.

In summary, we have discussed two possible observables, which 
allow to identify which of the decompositions of table \ref{Tab:TopoI} 
is realized, if a positive LNV observation is made at the LHC. 
The combination of both observables should be sufficient to 
identify the correct decomposition unambigously, apart from 
the two pairs: (A) 1-ii-a and 1-ii-b and (B) 1-i and 3-i (the latter 
only in the mass degenerate case), which can lead to very similar 
values in both observables.

\section{Summary}
\label{Sec:cncl}

In this paper we have compared the discovery potential of lepton
number violating signals at the LHC with the sensitivity of current
and future neutrinoless double beta decay experiments, assuming that
the decay rate is dominated by heavy ${\cal O}$ (TeV) particle
exchange. We have treated the first of two possible topologies
contributing to both processes which contains one fermion and two
bosons in the intermediate state, and concentrated on the case where
the fermion mass is always smaller than the scalar or vector
masses. The topology considered corresponds to 18 possible
decompositions including scalar, leptoquark and diquark
mechanisms. With the exception of some leptoquark mechanisms a
$0\nu\beta\beta$ decay signal corresponding to a half life in the
range $10^{26}-10^{27}$ yrs should imply a positive LNV signal at the
LHC, and vice versa, the non-observation of a positive signal at the
LHC would rule out a short-range mechanism for neutrinoless double
beta decay in most cases.  In summary the LHC search provides a
complementary and in many cases even superior option to search for
$\Delta L= 2$ lepton number violation for this short range case.

If $0\nu\beta\beta$ decay is triggered by light sub-eV scale Majorana
neutrinos, on the other hand, its LHC analogue will be unobservable.
In any case though an observation of either $0\nu\beta\beta$ decay or
its analogue at the LHC will prove also the light neutrinos to be
Majorana particles by virtue of the four loop contribution to the
neutrino mass generation according to the Schechter-Valle thorem
\cite{Schechter:1981bd,Hirsch:2006yk}. 
However, this 4-loop-induced Majorana mass while bestowing the light
neutrinos with Majorana-ness is too small to account for the mass
squared differences observed in neutrino oscillations
\cite{Duerr:2011zd}, implying
that differently generated masses, either of Dirac or of Majorana type
will be the dominant contributions for at least the heavier two mass
eigenstates.

Moreover, we have discussed two possibilities to discriminate
different contributions to the $0\nu\beta\beta$ decay rate by using
LHC observables: First, the charge asymmetry corresponding to the
ratio of positive like sign electron events and negative like sign
electron events, which reflects the larger abundance of $u$ quarks
compared to $d$ quarks in the most simple cases but becomes a more
complicated function of masses and couplings in the general case. For
large masses of the resonantly produced particles this asymmetry can
vary by up to 7 orders of magnitude.  And second, the resonance peaks
at the invariant mass distribution of the decay products of the heavy
particles produced on-shell.  The various resonance peaks depend on
the mass ordering of the intermediate particles and on the exact
decomposition and can then be used to identify the intermediate
particles triggering the decay.  Consequently, if an LNV signal at the
LHC would be found it should be possible to identify the dominant
contribution of $0\nu\beta\beta$ decay.

\medskip
\centerline{\bf Acknowledgements}

\medskip
We are grateful to Alfonso Zerwekh for useful discussions.
J.C.H. thanks the IFIC for hospitality during his stay.  This work was
supported by UNILHC PITN-GA-2009-237920 and by the Spanish MICINN
grants FPA2011-22975, MULTIDARK CSD2009-00064, by the Generalitat
Valenciana (Prometeo/2009/091), by Fondecyt (Chile) under grants
11121557, 1100582 and by CONICYT (Chile) project 791100017.  HP was
supported by DGF grant PA 803/6-1.

\setcounter{section}{0}
\def\theequation{\Alph{section}.\arabic{equation}}
\setcounter{equation}{0}

\section{Appendix A. Lagrangians }
\label{Sec:Lags}

Here we specify the Lagrangian terms used in our analysis. 

As was discussed in section \ref{Sec:Dec} there are two possible
topologies for the tree-level diagrams fig.~\ref{Fig:0nbbTopologies}
constructed of renormalizable interactions which contribute to
$0\nu\beta\beta$-decay and production of like-sign dileptons in
pp-collisions.  It is implied that gluons could be attached to any
colorful external or internal line of these diagrams. In the present
paper we focus on the Topology I corresponding to
fig.~\ref{Fig:0nbbTopologies}(a).  All the possible particles with
their SM assignments in the intermediate states of these diagrams are
listed in table \ref{Tab:TopoI} taken from ref. ~\cite{Bonnet:2012kh}.
These diagrams or their parts without the gluon insertions represent
mechanisms of $0\nu\beta\beta$-decay studied in the present paper.
Examples of Feynman diagrams are shown in figs. \ref{fig:Diags},
\ref{fig:LQprod}.

For our study it is sufficient to list renormalizable operators
corresponding to the vertices of these diagrams in the representation
with physical mass eigenstates after the electroweak symmetry
breaking. The SM gauge invariant representation in terms of the
electroweak interaction eigensates can be found in
Refs. ~\cite{Bonnet:2012kh,Han:2010rf}.  For the fields we adopt
notations
\begin{eqnarray}\label{notation-1}
F^{(n)}_{Q_{em}}
\end{eqnarray}
where $n$ is a dimension of the $SU(3)_{c}$ representation to which
belongs a field $F$ and $Q_{em}$ is its electric charge. We use $F =
S$ and $F = V_{\mu}$ for the scalar and vector fields respectively.

Below we specify the interactions of the scalar fields $S$. The
interactions of the vector fields can be readily derived from them by
substitution $S \rightarrow S^{\mu}$ with the same charge and
$SU_{3C}$ assignment and by simultaneous insertion of $\gamma_{\mu}$
to the coupled fermionic current.

We start with the interactions of the  scalar $S_{+1}^{(1,8)}$ fields participating in decompositions   1-i, 1-ii,  and 2-i, 2-ii from table  \ref{Tab:TopoI}. In our numerical analysis we use $S_{+1}$ for $S^{(1)}_{+1}$. These fields interact 
with quarks, charged leptons and fermions $\psi_{0}^{(1,8)}$ and $\psi_{5/3}^{(3)}, \psi_{4/3}^{(\bar{3})}$ according to 
the Lagrangian:
 \begin{eqnarray}
 \label{Lag-S1-psi}
 {\cal L}_{S_{+1}} &=& 
 \nonumber
 g^{(k)X}_{ud S_{+1}} \left( \bar u \  \hat{S}_{+1}^{(k)} P_X \ d\right)   +  
 g^{(k)X}_{e^{c} \psi_{0} S_{+1}} \left( \overline{e^{C}} \  P_X \  \psi^{(k)}_{0}\right)  S_{+1}^{(k)}   +
 g^{(k)X}_{e \psi_{0} S_{+1}}  \left( \overline{e} \  P_X \  \psi^{(k)}_{0}\right)   S_{+1}^{(k)\dagger} +\\
 &+&
 g^{X}_{u \psi_{5/3} S_{+1}}\  \left( \bar u \  P_X \   \psi^{(3)}_{5/3}  \right) S^{(1)}_{+1}    +
g^{X}_{d \psi_{4/3} S_{+1}}\  \left( \bar d \  P_X \   \psi^{(\bar{3})^{C}}_{4/3}  \right) S^{(1)}_{+1}    +  
\\
 \nonumber
&-& m^{(k)}_{\psi_{0}}\  \overline{\psi^{(k)}_{0}} \psi^{(k)}_{0} +   \mbox{h.c.} .
%
 \end{eqnarray}
The indices $X = L, P$ are independent in all the terms. The
generation indexes $i,j=1,2,3$ of the quarks $u, d$ and charged
leptons $e$ are suppressed.  We use the shorthand notations: $g^{(k)}
\hat{S}^{(k)} = g^{(1)}S^{(1)}$ {\bf I} $+ g^{(8)} S^{(8) A}
\lambda^{A}/2$ with the identity and Gell-Mann matrices in the color
space.  Also $g^{(k)} \psi^{(k)} S^{(k)} = g^{(1)} \psi^{(1)} S^{(1)}
+ g^{(8)} \psi^{(8)A} S^{(8)A}$ and $m^{(k)} \psi^{(k)} \psi^{(k)} =
m^{(1)} \psi^{(1)} \psi^{(1)} + m^{(8)} \psi^{(8)A} \psi^{(8)A} $.
For the $\psi_{0}$ field we study an economical case with only one
independent chiral component, so that in the 4-component notation it
is represented by a Majorana field satisfying $\psi_{0}^{C} =
\psi_{0}$. Thus, its mass $m_{\psi_{0}}$ in the last term of
eq. (\ref{Lag-S1-psi}) is a $\Delta L = 2$ Majorana mass. We do not
show the ordinary complex scalar mass term for $S_{+1}$ and the Dirac
ones for $\psi_{5/3}$, $\psi_{4/3}$.

The Majorana field cannot have definite lepton number, but for
convenience it can be assigned to the chiral projections of
$\psi_{0}$. One of the two options we choose is $L=1$ for $P_{L}
\psi_{0}$ and $L=-1$ for $P_{R} \psi_{0}$.  The fields $\psi_{5/3},
\psi_{4/3}$ and $S_{+1}$ have $L=0$. Baryon number $B$ conservation
requires an assignment $B=1/3$ for $\psi_{5/3}$, $B= -1/3$ for
$\psi_{4/3}$ and $B=0$ for $S_{+1},\ \psi_{0}$ .  As seen from
eq. (\ref{Lag-S1-psi}) there are two sources of LNV: the second
interaction term with the coupling $g^{L}_{e^{c}\psi S}$ and the
Majorana masses $m^{(1)}_{\psi_{0}}$ as well as $m^{(8)}_{\psi_{0}}$.

The scalar $SU_{3C}$ singlet field $S_{+2}$ appears in decompositions
1-ii, 3-ii and 3-iii of table \ref{Tab:TopoI}.  Its interactions are
given by:
\begin{eqnarray}\label{Lag-S2}
{\cal L}_{S_{+2}} &=& 
 g^{X}_{eeS_{+2}} \left( \overline{e^{C}} \  P_{X} \  e \right) S_{+2}   +  
 g^{X}_{d \psi_{5/3} S_{+2}} \left( \overline{\psi^{(3)}_{5/3}} \  P_X \  d \right)  S_{+2}   +\\
 \nonumber
&+& g^{X}_{u \psi_{4/3} S_{+2}} \left( \overline{u} \  P_X \  \psi^{(\bar{3})^{C}}_{4/3} \right)  S_{+2} + 
\mbox{h.c.}
\end{eqnarray}
Here the only LNV source is the first $\Delta L = 2$ term.

The diquarks $S^{DQ(3, \bar{6})}_{2/3}$ and $S^{DQ(\bar{3}, 6)}_{4/3}$
appear in decompositions 3-i, 3-ii, 3-iii, 4-ii and 5-ii of table
\ref{Tab:TopoI}.  These fields interact with quarks, charged leptons
and fermions $\psi^{(3)}_{5/3}, \psi^{(\bar{3})}_{4/3}$ and
$\psi^{(\bar{3}, 6)}_{1/3}$ in the following way:
 \begin{eqnarray}\label{Lagrangian-S6}
  {\cal L}_{DQ }&=& 
  g^{(6) X}_{uu S^{DQ}_{4/3}} \ (\bar{u} \ P_{X}\  \hat{S}^{DQ}_{4/3}\  u^{C} )  + 
  g^{(6) X}_{dd S^{DQ}_{2/3}} \ (\overline{d^{C}} \ P_{X}\  \hat{S}^{DQ}_{2/3} \ d )  +\\ 
  \nonumber
&+&  g^{(3) X}_{d_{i}d_{j} S^{DQ}_{2/3}} \  \epsilon^{IJK} (\overline{d^{C}}_{i I} \ P_{X}\ d_{j J}) S^{DQ(3)}_{2/3 K} +\\
\nonumber
&+& g^{(6) X}_{d\psi_{5/3} S^{DQ}_{4/3}} \ (\bar{d} \ P_{X}\   \hat{S}^{DQ}_{4/3}   \  \psi^{(3)^{C}}_{5/3} ) +
g^{(6) X}_{u\psi_{4/3} S^{DQ}_{2/3}} \ (\overline{\psi^{(\bar{3})}_{4/3}}  \     P_{X}\    \hat{S}^{DQ}_{2/3}   \   u) +\\
\nonumber
&+&  g^{(3) X}_{u\psi_{4/3} S^{DQ}_{2/3}} \  \epsilon^{IJK} (\overline{\psi^{(\bar{3})}_{4/3 I}}  \ P_{X}\  u_{J}) S^{DQ(3)}_{2/3 K} +\\
\nonumber
&+&  g^{X}_{e\psi_{1/3} S^{DQ}_{4/3}} \ (\overline{\psi^{(6)a}_{1/3}}  \ P_{X}\  e) \ S^{DQ (6)}_{4/3 a} + 
        g^{(6) X}_{e\psi_{1/3} S^{DQ}_{2/3}} \ (\overline{e^{C}}  \ P_{X}\  \psi^{(6)}_{1/3 a})\  S^{DQ(\bar{6}) a}_{2/3} + \\
\nonumber
&+&        g^{(3)X}_{e\psi_{1/3} S^{DQ}_{2/3}} \ 
(\overline{e^{C}}  \ P_{X}\  \psi^{(\bar{3}) I}_{1/3})\  
S^{DQ(3)}_{2/3\ I} \ .
 \end{eqnarray}

Here $I,J,K=1-3$ and $a=1-6$ are the color triplet and sextet indexes
respectively.  As before the generation indexes $i,j = 1,2,3$ of the
quarks $u, d$ and charged leptons $e$ are suppressed in all the terms
except for the third one which vanishes if $i=j$.  For convenience we
introduced notations $\hat{S}^{DQ}_{4/3} = S^{DQ (6)}_{4/3 a}
(T_{\bar{\bf 6}})^{a}_{IJ}$ and $\hat{S}^{DQ}_{2/3} = S^{DQ (\bar{6})
  a}_{2/3} (T_{\bf 6})_{a}^{IJ}$.  In the terms with these matrix
fields summation over the triplet indexes $I,J$ is implied.  The
symmetric $3\times 3$ matrices $T_{\bf 6}$ and $T_{\bar{\bf 6}}$ can
be found in ref.~\cite{Bonnet:2012kh}.  As seen from
eq. (\ref{Lagrangian-S6}) the sources of LNV in the diquark
interactions in are given by the last two $\Delta L = 2$ terms with
$\psi_{1/3}$ fields. We assign to $\psi_{1/3}^{(\bar{3}, 6)}$ a lepton
number $L=1$.

The leptoquark $SU_{3C}$ 3-plet fields $S^{LQ}_{2/3}$ and $
S^{LQ}_{-1/3}$ participate in decompositions 2, 4 and 5 of table
\ref{Tab:TopoI}.  Their interactions we write in the form:
\begin{eqnarray}\label{Lagrangian-LQ}
 %
  {\cal L}_{LQ} &=& 
  g^{X}_{eu S^{LQ}_{-1/3}} \  (\bar{u}^{I} \ P_X \ e^{C} ) \  S^{LQ}_{-1/3\ I} 
  + g^{X}_{ed S^{LQ}_{2/3}} \  (\bar{d}^{I} \ P_X\  e ) \  S^{LQ}_{2/3\ I} \\
  \nonumber
&+&  g^{(1) X}_{u \psi_{0} S^{LQ}_{2/3}} \  (\bar{u}^{I} \ P_X\  \psi_{0}^{(1)} ) \  S^{LQ}_{2/3\ I} +
 g^{(8) X}_{u \psi_{0} S^{LQ}_{2/3}} \  (\bar{u} \ P_X\  \hat\psi_{0} ) \  S^{LQ}_{2/3} +\\
 \nonumber
&+& g^{(1) X}_{d \psi_{0} S^{LQ}_{-1/3}} \  (\bar{d} \ P_X\  \hat\psi_{0}) \  S^{LQ}_{-1/3} +
 g^{(8) X}_{d \psi_{0} S^{LQ}_{-1/3}} \  (\bar{d} \ P_X\  \hat\psi_{0}) \  S^{LQ}_{-1/3} +\\
 \nonumber
&+&  g^{(3) X}_{d \psi_{1/3} S^{LQ}_{2/3}} \  \epsilon_{IJK} (\bar{d}^{I} \ P_X\  \psi_{1/3}^{(\bar{3})J}) \  S^{LQ, K^{\dagger}}_{2/3} +
g^{(6) X}_{d \psi_{1/3} S^{LQ}_{2/3}} \  (\bar{d} \ P_X\  \hat\psi_{1/3}) \  S^{LQ^{\dagger}}_{2/3} \\
 \nonumber
 &+&  g^{(3) X}_{u \psi_{1/3} S^{LQ}_{-1/3}} \  \epsilon_{IJK} (\bar{u}^{I} \ P_X\  \psi_{1/3}^{(\bar{3}) J}) \  S^{LQ,K^{\dagger}}_{-1/3}+
g^{(6) X}_{u \psi_{1/3} S^{LQ}_{-1/3}} \  (\bar{u} \ P_X\  \hat\psi_{1/3}) \  S^{LQ^{\dagger}}_{-1/3}\\
 \nonumber
&+&  g^{X}_{e \psi_{4/3} S^{LQ}_{-1/3}} \  (\overline{e^{C}} \ P_X\  \psi_{4/3}^{(\bar{3}) I}) \  S^{LQ}_{-1/3\ I} +
g^{X}_{e \psi_{5/3} S^{LQ}_{2/3}} \  (\overline{e^{C}} \ P_X\  \psi_{5/3\ I}^{(3)}) \ S^{LQ, I^{\dagger}}_{2/3} \ .
\end{eqnarray}
As before we introduce a short-hand notation $\hat\psi_{1/3} =
\psi^{(6)}_{1/3\ a} (T_{\bar{\bf 6}})^{a}_{IJ}$. Here $I,J= 1,2,3$ are
the color triplet indexes.  We adopt the following assignment of
lepton $L$ and baryon numbers to the leptoquarks: $L=1, \ B=1/3$ for
$S^{LQ}_{-1/3}$ and $L=-1, \ B=1/3$ for $S^{LQ}_{2/3}$.  Checking the
total lepton number of each term in eq. (\ref{Lagrangian-LQ}) one
finds that the terms in the 2nd line with chirality $X=R$, in the 3rd
line with $X=L$, in the 4th and the last lines with any $X$ break 
lepton number in two units.

The following comments on the structure of the $\Delta L=2$ amplitude
is in order.  For the analysis of $0\nu\beta\beta$-decay we introduced
in eq. (\ref{eq:meff}) an effective masses $M_{eff}$ and an effective
couplings $g_{eff}$.  The quantities $M_{eff}^{5}$ and $g_{eff}^{4}$
represent respectively the products of the particle masses originating
from their propagators and the products of four couplings, $g_{i}$, of
those operators from eqs. (\ref{Lag-S1-psi})-(\ref{Lagrangian-LQ})
which participate in the decomposition in question.  Let us specify
possible characteristic cases for combinations of these masses and
couplings in $0\nu\beta\beta$ amplitude.  Schematically one can
distinguish the following cases:
\begin{eqnarray}\label{Rules-Ampl}
{\cal A}(0\nu\beta\beta) &\sim& 
g_{1} g_{2\psi_{0}}^{X} g_{3\psi_{0}}^{X} g_{4} \frac{m_{\psi_{0}}}{m^{2}_{S_{1}} m^{2}_{S_{2}} m^{2}_{\psi_{0}}}\ , 
\\
\nonumber
&\sim& \g_{1} g_{2\psi_{Q}}^{X} g_{3\psi_{Q}}^{X} g_{4} \frac{m_{\psi_{Q}}}{m^{2}_{S_{1}} m^{2}_{S_{2}} m^{2}_{\psi_{Q}}}\ , \ \ \ 
\g_{1} g_{2\psi}^{L} g_{3\psi}^{R} g_{4} \frac{\langle \gamma_{\mu} q^{\mu}\rangle}{m^{2}_{S_{1}} m^{2}_{S_{2}} m^{2}_{\psi}}\ , \\
\nonumber
&\sim&  
g_{1} \g_{2\psi_{0}}^{X} \g_{3\psi_{0}}^{X} g_{4} \frac{m_{\psi_{0}}}{m^{2}_{S_{1}} m^{2}_{S_{2}} m^{2}_{\psi_{0}}}  .
\end{eqnarray} 
Here, $X = L,R$ and $g_{i\psi}$ are the couplings of the operators
involving $\psi$-field. These fields with nonzero charge $Q$ are
denoted as $\psi_{Q}$. Without this index they can be both charged
$\psi_{Q}$ and neutral $\psi_{0}$.  The masses $m_{\psi_{Q}}$ of the
$\psi_{Q}$ fields are of Dirac $\Delta L= 0$ type while in the case of
the $\psi_{0}$ fields their masses $m_{\psi_{0}}$ are of Mjorana
$\Delta L=2$ type.  By $\g_{i}$ we denote the coupling of the $\Delta
L = 2$ operators. In the case of only one slash, as in the second
line, it may take place in any of the four couplings, while the two
slashed couplings can only be of the $\g_{i\psi_{0}}$-type as in the
last line.  In the combination given in the first line the $\Delta L
=2$ is brought by the Majorana mass $m_{\psi_{0}}$ while in both cases
of the second line it is due to a single $\g_{i}$ coupling.  The
combination of the last line put together three sources of the $\Delta
L = 2$ in a total $\Delta L = 2$.  Note that the expressions in
eq. (\ref{Rules-Ampl}) imply that the masses of the intermediate
particles $m_{i} \gg |{\bf q}|$ where ${\bf q}$ are their momenta
whose mean value is about $\sim 100$ MeV.  The numerator of the third
combination $\langle \gamma_{\mu} q^{\mu} \rangle$ implies inception
of $\gamma_{\mu}$ in between of the two electron or two quark
bispinors depending on the considered decomposition.  It is of the
order of ${\bf q}\sim$ 100 MeV and, therefore, the third term
corresponding to the LR chirality structure is suppressed in
comparison with the remaining LL or RR terms by a factor of ${\bf
  q}/m_{\psi}$.  Thus, among all the possible cases specified in
(\ref{Rules-Ampl}) survive only those with LL or RR chiralities
leading to
\begin{eqnarray}\label{Summary}
{\cal A}(0\nu\beta\beta) \sim \frac{g_{eff}^{4}}{M_{eff}^{5}}
\end{eqnarray}
in terms of the effective quantities introduced in eq. (\ref{eq:meff}).  
The decompositions leading to the third term in eq. (\ref{Rules-Ampl})
are very weakly constrained by $\znbb$ decay experiments.  However
they can be probed at the LHC in the way we discussed in the main
text.

\end{document}